\documentclass{article}[12pt]
\usepackage{amsfonts}

\newbox\strutbox
\setbox\strutbox=\hbox{\char"B}
\def\strut{\relax\ifmmode\copy\strutbox\else\unhcopy\strutbox\fi}
\def\ialign{\everycr{}\tabskip0pt\halign}
\def\eqalign#1{\null \,\vcenter {\openup\jot \mathsurround 0pt
    \ialign{\strut \hfil$\displaystyle{##}$&$\displaystyle
      {{}##}$\hfil\crcr#1\crcr}}\,}

\def\eqalignno#1{\tabskip 0pt plus 1 fill \halign to\displaywidth{\hfil$\tabskip0pt\everycr{}\displaystyle{##}$\tabskip 0pt &$\tabskip0pt\everycr{}\displaystyle {{}##}$\hfil \tabskip 0pt plus 1 fill&\llap {$\tabskip0pt\everycr{} ##$}\tabskip 0pt \crcr #1\crcr }}

\begin{document}
\title{Spin-2 Amplitudes in Black-Hole Evaporation}
\author{A.N.St.J.Farley and P.D.D'Eath\protect\footnote{Department of Applied Mathematics and Theoretical Physics,
Centre for Mathematical Sciences,
University of Cambridge, Wilberforce Road, Cambridge CB3 0WA,
United Kingdom}}

\maketitle

\begin{abstract}
Here, quantum amplitudes for $s=2$ linearised gravitational-wave
perturbations of a spherically-symmetric Einstein/massless-scalar 
background, describing gravitational collapse to a black hole, are 
treated by analogy with the previous treatment of $s=1$ linearised 
Maxwell-field perturbations.  As with the spin-1 case, the spin-2 
perturbations split into parts with odd and even parity.  Their 
detailed angular behaviour is analysed, as well as their behaviour
under infinitesimal coordinate transformations and their (linearised
vacuum Einstein) field equations.  In general, we work in the
Regge-Wheeler gauge, except that, at a certain point, it becomes
necessary to make a gauge transformation to an asymptotically-flat
gauge, such that the metric perturbations have the expected fall-off
behaviour at large radii.  As with the $s=1$ treatment, so in this
$s=2$ case we isolate suitable 'coordinate' variables which can be
taken as boundary data on a final space-like hypersurface $\Sigma_F{\,}$.
(For simplicity of exposition, we take the data on the initial surface 
$\Sigma_I$ to be exactly spherically-symmetric.)  The (large)
Lorentzian proper-time interval between $\Sigma_I$ and $\Sigma_F{\,}$,
measured at spatial infinity, is denoted by $T{\,}$.  We then consider 
the classical boundary-value problem and calculate the 
second-variation classical Lorentzian action 
$S^{(2)}_{\rm class}{\;}$, on the assumption that the time interval
$T$ has been rotated into the complex: 
$T\rightarrow{\mid}T{\mid}\exp(-i\theta){\,}$, for 
$0<\theta\leq\pi/2{\,}$.  This complexified classical boundary-value 
problem is expected to be well-posed, in contrast to the 
boundary-value problem in the Lorentzian-signature case $(\theta =0)$, 
which is badly posed, since it refers to hyperbolic or wave-like field 
equations.  Following Feynman, we recover the Lorentzian quantum 
amplitude by taking the limit as $\theta\rightarrow 0_{+}$ of the 
semi-classical amplitude ${\,}\exp(iS^{(2)}_{\rm class}){\,}$.  
The boundary data for $s=2$ involve the magnetic part of the Weyl
tensor, just as the data for $s=1$ involve the usual (Maxwell)
magnetic field.  These relations are also investigated, using 
2-component spinor language, in terms of the Maxwell field strength 
$\phi_{AB}{\,}={\,}\phi_{(AB)}$ and the Weyl spinor 
${\,}\Psi_{ABCD}{\,}={\,}\Psi_{(ABCD)}{\,}$.
\end{abstract} 
\begin{section}{Introduction}
In this paper, we apply to the $s=2$ gravitational-wave perturbations
a treatment analogous to that given for $s=1$ Maxwell (photon) 
perturbations in [1].  Just as in [1] for $s=1{\,}$, the $s=2$
graviton perturbations about a spherically-symmetric background
Einstein/massless-scalar model describing gravitational collapse, may
be analysed in terms of suitable angular harmonics, having either odd
or even parity.  The classical (second-variation) Lorentzian action
$S^{(2)}_{\rm class}$ is derived, as a functional of suitably chosen
boundary data on the final space-like hypersurface $\Sigma_F$ at a
very late proper time $T{\,}$, measured at spatial infinity, in each 
of the odd and even cases.  For simplicity of exposition, we assume that
the $s=2$ boundary data on the initial pre-collapse hypersurface
$\Sigma_I{\,}$, at an early time $t=0$ (say), are zero.  This would
correspond to the initial intrinsic 3-metric $h_{ij}$ on $\Sigma_I$
being exactly spherically-symmetric.

As with spin-0 scalar perturbations [2,3] and $s=1$ Maxwell 
perturbations [1], one can deduce the Lorentzian quantum amplitude to 
go from prescribed initial data on $\Sigma_I$ (spherically-symmetric, 
in the above instance) to prescribed final perturbative data on 
$\Sigma_F{\,}$, from the study of the perturbative (second-variation) 
Lorentzian action $S_{\rm class}$ as a function of complexified 
time-at-spatial-infinity $T{\,}$.  As before, one first rotates 
$T\rightarrow{\mid}T{\mid}\exp(-i\theta)$ into the complex, for 
$0<\theta\leq\pi/2{\,}$.  Then the classical (Dirichlet or Neumann) 
linearised boundary-value problem is expected to become well-posed.  
The classical linearised solution for fixed (real) boundary data but 
variable complex $T$ (with $\theta > 0$) becomes complex-valued.  
Similarly, the second-variation classical Lorentzian action 
$S_{\rm class}$ becomes a complex-valued functional of the (real) 
boundary data, and a function of the complex variable $T{\,}$,
provided ${\rm Im}(T)<0{\,}$.  Following Feynman [4], one recovers the
(complex) quantum amplitude to go from the prescribed initial data on
$\Sigma_I$ to final data on $\Sigma_F{\,}$, by considering the
semi-classical amplitude, proportional to $\exp(iS_{\rm class}){\,}$,
and then taking the limit as $\theta\rightarrow 0_{+}{\;}$. The quantum
amplitude in the case $0<\theta\leq\pi/2$ contains also loop
corrections, multiplying ${\,}\exp(iS_{\rm class}){\,}$.  For quantum
amplitudes to be meaningful, one expects that the theory should be 
invariant under local supersymmetry [2].  In the locally-supersymmetric 
case, for supergravity coupled to supermatter [12,13], the loop 
corrections will have a very small effect, provided that the 
frequencies involved in the boundary data are small compared with 
the Planck scale.

In Sec.2, a more unified treatment of the angular harmonics appearing
in [1] for $s=1{\,}$, and in the present paper for $s=2{\,}$, is
outlined in terms of vector and tensor spherical harmonics [5,6].  
The detailed angular decomposition of the metric perturbations for
odd-parity $s=2$ harmonics is given in Sec.3; this is simplified by
use of the Regge-Wheeler (RW) gauge [7], and the odd-parity
linearised vacuum Einstein equations are given.  Sec.4 begins the
process of computing the classical action $S^{(2)}_{\rm class}$ for
odd-parity metric perturbations.  The boundary conditions on the
odd-parity $s=2$ perturbations are discussed in Sec.5.  Both the odd-
and even-parity metric perturbations, obeying the linearised vacuum
Einstein equations, inevitably grow at a rate $O(r)$ at large radius
$(r\rightarrow\infty)$, compared to an asymptotically-flat background
metric, when viewed in RW gauge.  For odd-parity metric perturbations,
in Sec.6 a gauge transformation is given to a new asymptotically-flat
(AF) gauge, such that the metric perturbations in the AF gauge die
away at large $r{\,}$, as expected on physical grounds.  This leads 
to a convenient choice of variables with which to describe data on the
final surface $\Sigma_F{\,}$, for odd-parity metric perturbations.  An
explicit expression for the odd-parity perturbative (second-variation)
part of the classical action $S_{\rm class}$ is given in Eq.(6.17).
From this, as described above, one can compute the classical 
Lorentzian quantum amplitude.

In Sec.7, we turn to the even-parity metric perturbations.  The
detailed angular decompositions are given there, together with a
treatment of even-parity gauge transformations and of the linearised
vacuum Einstein field equations in RW gauge.  The corresponding
classical action functional is introduced in Sec.8.  As with the
odd-parity metric perturbations in Sec.5, so in Sec.9 a convenient
description of boundary data for even-parity metric perturbations is
given, so that the second-variation classical action $S_{\rm class}$
can be given explicitly in terms of the data, in Eq.(9.7). From this
expression for $S_{\rm class}{\,}$, one can again deduce the
Lorentzian quantum amplitude from taking the limit 
${\,}\theta\rightarrow 0_{+}{\;}$, by analogy with the spin-0 
and spin-1 cases [1,2].

In Sec.10, we note that the boundary data above, suitable as arguments
of the wave functional $\Psi$, involve specifying (odd- or even-parity 
parts of) the magnetic part $H_{ik}$ of the Weyl tensor [8-11]. 
Similarly, in Sec.6 of [1], we considered, in 2-component spinor 
language, the analogous $s=1$ Maxwell boundary data, involving the 
magnetic field $B_i{\,}$.  For $s=1{\,}$, knowledge of (real) $B_i$ 
is equivalent to knowledge of 3 of the 6 real components of the 
symmetric Maxwell field-strength spinor
$\phi_{AB}{\,}={\,}\phi_{(AB)}{\,}$ [8,9].  Correspondingly, 
in the $s=2$ case of the present paper, knowledge of the 5 real, 
symmetric, trace-free components of $H_{ik}$ at a point is equivalent 
to knowledge of half of the 10 real components of the
totally-symmetric Weyl spinor 
$\Psi_{ABCD}{\,}={\,}\Psi_{(ABCD)}{\,}$ [8,9,12].  
Sec.11 contains a brief conclusion.
\end{section}

\begin{section}{Vector and tensor spherical harmonics}
A more unified treatment of the angular harmonics appearing 
for $s=1$ (Maxwell) and $s=2$ (graviton) perturbations of a
spherically-symmetric background can be given in terms of vector 
and tensor harmonics [5,6].  In [2,3], we expanded the $s=0$ 
(massless-scalar) perturbations in terms of scalar spherical harmonics 
$Y_{\ell m}(\theta,\phi)$, which have even parity.  Vector and tensor 
spherical harmonics, however, can have odd as well as even parity.

Any vector field in a spherically-symmetric background, such as the
classical $s=1$ (photon) solutions appearing in [1], can be expanded 
in terms of vector spherical harmonics on the unit 2-sphere.  
Correspondingly, angular vector and tensor indices are raised 
and lowered with the metric $\hat\gamma_{ab}{\,}$, given by 
$$\hat\gamma_{\theta\theta}{\;}{\,}={\;}{\,}1{\;},
{\qquad}\hat\gamma_{\phi\phi}{\;}{\,}
={\;}{\,}\sin^{2}\theta{\;},
{\qquad}\hat\gamma_{\theta\phi}{\;}{\,}={\;}{\,}\hat\gamma_{\phi\theta}{\;}{\,}
={\;}{\,}0{\quad}.\eqno(2.1)$$
\noindent
The even-parity vector harmonics [5,6] have angular components
$$(\Psi _{\ell m})_a{\;}{\,}={\;}{\,}\partial_{a}Y_{\ell m}{\quad},\eqno(2.2)$$
\noindent 
where ${\,}a=(\theta,\phi){\,}$.  The odd-parity vector harmonics are
$$(\Phi_{\ell m})_{a}{\;}{\,} 
={\;}{\,}\epsilon_{a}^{~~b}{\,}(\partial_{b}Y_{\ell m}){\quad}.\eqno(2.3)$$
\noindent
Here, $\epsilon_{a}^{~~b}$ denotes the tensor with respect to angular 
indices $(a =\theta,\phi{\;};{\;}b=\theta,\phi)$, such that the 
lowered version ${\,}\epsilon_{ab}=-{\,}\epsilon_{ba}{\,}$ 
is anti-symmetric, with
${\,}\epsilon_{01}=\bigl({\hat\gamma}\bigr)^{{1}\over{2}}=\sin\theta{\;}$,
where ${\,}{\hat\gamma}={\rm det}\bigl(\hat \gamma_{ab}\bigr){\,}$.  Thus,
$$\epsilon_{\theta}^{~~\phi}{\;}{\,}={\;}{\,}{{-1}\over{\sin\theta}}{\quad},
{\qquad}\epsilon_{\phi}^{~~\theta}{\;}{\,}={\;}{\,}\sin\theta{\quad},
{\qquad}\epsilon_{\phi}^{~~\phi}{\;}{\,}={\;}{\,}\epsilon{_\theta}^{~\theta}{\;}{\,} 
={\;}{\,}0{\quad}.\eqno(2.4)$$
\noindent
The forms of the angular harmonics appearing in the $s=1$ photon
calculations of [1] can be deduced from these
vector-spherical-harmonic expressions.

Analogously, any rank-2 tensor field such as a linearised (graviton) 
$s=2$ classical solution, to be treated in this paper, can be expanded 
in terms of tensor spherical harmonics.  The even-parity harmonics are
$$(\Psi_{\ell m})_{ab}{\;}{\,}={\;}{\,}Y_{\ell m\mid ab}{\quad}, 
{\qquad}(\Phi_{\ell m})_{ab}{\;}{\,}={\;}{\,}\hat\gamma_{ab}{\;}Y_{\ell m}{\quad},
\eqno(2.5)$$
\noindent
where a bar $\mid$ denotes a covariant derivative with respect to
the metric $\hat\gamma_{ab}{\,}$.  The odd-parity harmonics are
$$(\chi_{\ell m})_{ab}{\;}{\,}
={\;}{\,}{{1}\over{2}}\biggl[\epsilon_{a}^{~~c}{\;}\Bigl(\Psi_{\ell m}\Bigr)_{cb}{\,} 
+{\;}\epsilon_{b}^{~~c}{\;}\Bigl(\Psi_{\ell m}\Bigr)_{ca}\biggr]{\quad}.
\eqno(2.6)$$
\indent
Our boundary-value problem, as posed in [2,3], involves specifying on
the final space-like hypersurface ${\,}\Sigma_F{\,}$ the spatial
components ${\,}h^{(1)}_{ij}(x){\,}$, for ${\,}i,j = 1,2,3{\,}$, of the real
perturbations ${\,}h^{(1)}_{\mu\nu}(x){\,}$ of the 4-metric 
${\,}(\mu,\nu =0,1,2,3)$.  We shall construct a basis of tensor spherical 
harmonics with which to expand the angular dependence of
$h^{(1)}_{ij}{\,}$.  In general, we make a multipole decomposition for 
real metric perturbations, of the form:
$$h^{(1)}_{ij}(x){\;}{\,}
={\;}{\,}\sum^{\infty}_{\ell =2}{\;}\sum^{\ell}_{m=-\ell}{\;}
\biggl[\bigl(h^{(-)}_{ij}\bigr)_{\ell m}(x){\,} 
+{\;}\bigl(h^{(+)}_{ij}\bigr)_{\ell m}(x)\biggr]{\quad},\eqno(2.7)$$
\noindent
where $-$ and $+$ denote odd- and even-parity contributions,
respectively.  The limit $\ell =2$ in the summation over $\ell$ 
will be commented on in the following Sec.3.  
\end{section}

\begin{section}{Odd-parity perturbations}
In [1], $s=1$ (Maxwell) perturbations of spherically-symmetric 
gravitational backgrounds were treated in the Regge-Wheeler (RW) 
formalism [7], and split naturally into odd and even type, 
according to their behaviour under parity inversion:
$\theta \rightarrow\;(\pi-\theta),\; \phi\rightarrow\;(\pi +\phi).$  
The even 'electric-type' perturbations have parity 
$\pi = (-1)^{\ell}$, while the odd 'magnetic-type' perturbations 
have parity $\pi =(-1)^{\ell + 1}$.  The analogous (orthogonal) 
decomposition also holds for the $s=2$ 
gravitational-wave perturbations.

In the $s=1$ Maxwell case, the lowest $\ell =0$ mode does not
propagate:  in the electric case, it corresponds to the addition of a
small charge to the black hole, to turn a Schwarzschild solution into
a Reissner-Nordstr\"om solution with charge $Q{\ll}M{\,}$.
Correspondingly, in the $s=2$ case of gravitational perturbations, the
multipoles with $\ell<2$ are non-radiatable.  For example, the
even-parity gravitational perturbations with $\ell = 0$ correspond to
a small static charge in the mass, while the $\ell = 0$ odd-parity
perturbation is identically zero.  For $\ell =1$, the odd-parity
(dipole) gravitational perturbations must be stationary [13], and
even-parity dipole perturbations correspond to a coordinate
displacement of the origin [14]  and can be removed by a gauge
transformation.  For a general spin $s=0, 1, 2,$
perturbations with $\ell<{\mid}s{\mid}$ relate to total conserved
quantities in the system.  In the present $s=2$ gravitational-wave
case, we consider only the propagating $\ell = 2$ (quadrupole) and
higher-$\ell$ modes.

In this Section and the following Secs.4,5, we restrict attention to
odd-parity gravitational-wave perturbations.  Following Moncrief [15],
we write
$$\Bigl(h^{(-)}_{ij}\Bigr)_{\ell m}(x)\; 
= \; h^{(-)}_{1\ell m}(t,r)
\biggl[(e_{1})_{ij}\biggr]_{\ell m}\;
+\; h^{(-)}_{2\ell m}(t,r)\biggl[(e_{2})_{ij}\biggr]_{\ell m}.\eqno(3.1)$$
\noindent
(N.B. : one should not confuse the subscripts $1,2$ here with spin
subscripts.)  The non-zero components of the symmetric tensor fields 
$[(e_{1,2})_{ij}]_{\ell m}$ are defined by
$$((e_1)_{r\theta})_{\ell m}{\;}
={\;}-{\,}({\partial}_{\phi}Y_{\ell m})/(\sin\theta),\eqno(3.2)$$
$$((e_1)_{r\phi})_{\ell m}{\;}
={\;}(\sin\theta)({\partial}_{\theta}Y_{\ell m}),\eqno(3.3)$$
$$((e_2)_{\theta\theta})_{\ell m}{\;}
={\;}(\sin\theta)^{-2}{\,}
\bigl((\sin\theta){\partial}^{2}_{\theta\phi}{\,}
-{\,}(\cos\theta){\partial}_{\phi}\bigr) Y_{\ell m},\eqno(3.4)$$
$$((e_2)_{\theta\phi})_{\ell m}{\;}
={\;}{{1}\over{2}}{\,}
\bigl((\sin\theta)^{-1}{\,}{\partial}^{2}_{\phi}{\,}
-{\,}(\cos\theta){\,}{\partial}_{\theta}{\,}
-{\,}(\sin\theta){\,}{\partial}^{2}_{\theta}\bigr){\,}Y_{\ell m},
\eqno(3.5)$$
$$((e_2)_{\phi\phi})_{\ell m}{\;}
={\;}\bigl((\cos\theta){\partial}_{\theta}{\,}
-{\,}(\sin\theta){\,}{\partial}^{2}_{\theta\phi}\bigr){\,}Y_{\ell m}.
\eqno(3.6)$$

This basis is normalised according to:
$$\eqalignno{\int\;d\;\Omega\biggl[(e_{1})^{ij}\biggr]_{\ell m}\;
\biggl[(e_{2})_{ij}\biggr]^{*}_{\ell' m'}\; 
&= \; 0,&(3.7)\cr
\int\;d\;\Omega\biggl[(e_{1})^{ij}\biggr]_{\ell m}\;
\biggl[(e_{1})_{ij}\biggr]^{*}_{\ell' m'}\; 
&=\;{{2e^{-a}}\over{r^{2}}}\ell(\ell +1)
\delta_{\ell\ell'}\;\delta_{mm'},&(3.8)\cr
\int\;d\;\Omega\biggl[(e_{2})^{ij}\biggr]_{\ell m}\;
\biggl[(e_{2})_{ij}\biggr]^{*}_{\ell' m'}\; 
&=\;{{\ell(\ell +1)(\ell +2)(\ell -1)}\over{2r^{4}}}
\delta_{\ell\ell'}\;\delta_{mm'},&(3.9)\cr}$$
\noindent
where 
$\int d\Omega(~~) 
=\int^{2\pi}_{0}\;d\phi\;\int^{\pi}_{0}\;d\theta\;\sin \theta(~~){\,}$,
and where these indices are raised and lowered using the background
3-metric $\gamma^{ij}{\,},\gamma_{ij}{\,}$.  Note that both
$\bigl[(e_{1})_{ij}\bigr]_{\ell m}$ and
$\bigl[(e_{2})_{ij}\bigr]_{\ell m}$ are traceless.

In the standard $(3+1)$ decomposition for the gravitational field [16],
the 4-metric $g_{\mu\nu}$ is decomposed into the spatial 3-metric
${\,}h_{ij}{\,}={\,}g_{ij}{\,}$ on a hypersurface
$\{x^{0}={\rm const.}\}$, together with the lapse function $N$ and
the shift vector field $N^i{\,}$.  For odd-parity perturbations 
of the lapse, one has
$$N^{(1)(-)}\;=\; 0,\eqno(3.10)$$
\noindent
while the odd-parity shift takes the form
$$\Bigl[N_{i}^{~(-)}\Bigr]_{\ell m}\; 
=\; h^{(-)}_{0\ell m}(t,r)
\biggl[0,{\,}-{1\over{(\sin\theta)}}(\partial_{\phi}Y_{\ell m}),
(\sin\theta)(\partial_{\theta}Y_{\ell m})\biggr].\eqno(3.11)$$
\noindent
For a real 4-metric $g_{\mu\nu}{\,}$, both $h^{(1)}_{ij}$ and
$N^{(1)(-)}_{i}$ are real, and one has
$$h^{(-)*}_{0,1,2\ell m}\; 
= (-1)^{m}\; h^{(-)}_{0,1,2\ell,-m}{\quad}.\eqno(3.12)$$

In the Hamiltonian formulation of general relativity, the momentum 
$\pi^{ij}$ conjugate to the 'coordinate' $h_{ij}$ is a symmetric 
spatial tensor density.  As with the 3-metric $h_{ij}$ above 
[Eq.(3.1)], the linearised perturbations of $\pi_{ij}$ can be 
decomposed into multipoles with odd or even parity:
$$\pi^{(1)}_{ij}(x)\; 
=\;\sum^{\infty}_{\ell = 2}\sum^{\ell}_{m=-\ell}
\biggl[(\pi^{(-)}_{ij})_{\ell m}(x) \;
+\;(\pi^{(+)}_{ij})_{\ell m} (x)\biggr].\eqno(3.13)$$
\noindent
Restricting attention at present to the odd modes, one has
$$(\pi^{(-)}_{ij})_{\ell m}\; 
=\;(^{(3)}\gamma)^{1 \over 2}\Biggl\{p_{1\ell m}(t,r)
\Bigl[(e_{1})_{ij}\Bigr]_{\ell m}\; 
+\; p_{2 \ell m}(t,r)
\Bigl[(e_{2})_{ij}\Bigr]_{\ell m}\Biggr\},\eqno(3.14)$$
\noindent
where $[(e_{1})_{ij}]_{\ell m}$ and $[(e_{2})_{ij}]_{\ell m}$ 
are given above.  One finds that
$$\eqalignno{p_{1\ell m}(t,r)\; 
&=\;{1\over {2N^{(0)}}} \;
\biggl[\partial_{t}\;h^{(-)}_{1\ell m}\;
-r^{2}\;\partial_{r}\biggl({{h^{(-)}_{0\ell m}\over{r^{2}}}}\biggr)\biggr],
&(3.15)\cr
p_{2\ell m}(t,r)\; 
&=\; {{1}\over{2N^{(0)}}}
\biggl[\partial_{t}\;h^{(-)}_{2\ell m} \;
+2h^{(-)}_{0\ell m}\biggr].&(3.16)\cr}$$

One can typically simplify the form of the perturbations by a
gauge transformation (linearised coordinate transformation) 
in a neighbourhood of the final space-like hypersurface 
$\Sigma_F{\,}$.  Suppose that the infinitesimal transformation 
is along a vector field $\xi^{\mu}{\,}$.  Then the metric 
perturbations transform infinitesimally by
$$g_{\mu\nu} \rightarrow g_{\mu\nu}\; 
-\;\nabla_{\mu}\;\xi_{\nu}\;-\;\nabla_{\nu}\;\xi_{\mu}\;.\eqno(3.17)$$ 
\noindent
For odd perturbations, in the notation of Eq.(2.2), consider the
infinitesimal vector field $\xi^{(-)\mu}{\,}$, with components [7]  
given by
$$\eqalign{\Bigl(\xi^{(-)t}\Bigr)_{\ell m}\; 
&=\;0,\quad\Bigl(\xi^{(-)r}\Bigr)_{\ell m}\;
=\;0,\cr
\Bigl(\xi^{(-)a}\Bigr)_{\ell m}\; 
&=\;{{\Lambda _{\ell m}(t,r)}\over{r^{2}}}
(\Phi_{\ell m})^{a}\;.\cr}\eqno(3.18)$$
\noindent
The resulting 'gauge transformation' is summarised by
$$\eqalignno{h^{(-)'}_{0\ell m}\; 
&=\; h^{(-)}_{0\ell m}\; 
-\;\partial_{t}\Lambda_{\ell m}\;,&(3.19)\cr
h^{(-)'}_{1\ell m}\;
&=\;h^{(-)}_{1\ell m}\; 
-\;\partial_{r}\;\Lambda_{\ell m}\;
+\;{{2 \Lambda_{\ell m}}\over r}\;,&(3.20)\cr
h^{(-)'}_{2 \ell m}\; 
&=\; h^{(-)}_{2\ell m}\; 
+\;2\;\Lambda _{\ell m}\;\;.&(3.21)\cr}$$

We have here neglected time-derivatives of the metric components: 
we are assuming that an Ansatz for the gauge functions 
$\Lambda_{\ell m}(t,r)$ based on separation of variables will be 
valid, involving frequencies which satisfy the adiabatic approximation 
[2,3] described below.  In the Regge-Wheeler gauge [7], we set 
$h^{(-)RW}_{0\ell m}\;
=\; h^{(-)'}_{0\ell m}$ 
and $h^{(-)RW}_{1\ell m}=\;h^{(-)'}_{1\ell m}$, 
as in Eqs.(3.19,20), but require also
$$h^{(-)RW}_{2 \ell m}\; 
=0\;
=\; h^{(-)}_{2\ell m}\;
+2\;\Lambda_{\ell m}.\eqno(3.22)$$ 
\noindent
For each $\ell{\,}$, one can obtain solutions for arbitrary $m$ 
by rotation from the case $m=0{\,}$.  Note also that the above 
equations show how the RW perturbations can be uniquely 
recovered from the perturbations in an arbitrary gauge.

Since odd-and even-parity perturbations decouple, the odd-parity 
field equations in the RW gauge are obtained by substituting 
Eq.(2.7), together with the equation $N^{(-)}_{i} =\;h^{(-)}_{ti}$ 
for the linearised shift vector, into the source-free linearised 
Einstein field equations [16], as given in [17].  The (Lorentzian)
spherically-symmetric background metric is taken, as in [2,3], to be:
$$ds^{2}\; 
=\; -e^{b(t,r)}\;dt^{2}\; + e^{a(t,r)}\; dr^{2}
+\;r^{2}(d\theta^{2} +\sin^{2}\theta\;d\phi^{2}).\eqno(3.23)$$
\noindent
Then the odd-parity linearised field equations, taking respectively
the $(t\phi), (r\phi)$ and $(\theta\phi)$ components, read:
$$(\partial_{r})^{2}\;h^{(-)RW}_{0\ell m}\;
-\;\partial_{t}\partial_{r}\; h^{(-)RW}_{1\ell m}
-{{2}\over{r}}\;\partial_{t}h^{(-)RW}_{1\ell m}\;
+\;F_{1 \ell}(t,r)\; h^{(-)RW}_{0\ell m}
=\;0,\eqno(3.24)$$ 
\noindent
and 
$$\eqalign{&(\partial_{t})^{2}h^{(-)RW}_{1\ell m}
-\partial_{t}\;\partial_{r}\; h^{(-)RW}_{0\ell m}
+{{2}\over{r}}\;\partial_{t}\;h^{(-)RW}_{0\ell m}
+{{1}\over{2}}(\dot a \;+\dot b)
\biggl[\partial_{r}\;h^{(-)RW}_{0\ell m}
-\partial_{t}\;h^{(-)RW}_{1\ell m}\biggr]\cr
&-\biggl[{1 \over r}(\dot a+\dot b)
+{{1}\over{2}}b'(\dot a -\dot b)\biggr]h^{(-)RW}_{0\ell m}
-F_{2\ell}(t,r)\;h^{(-)RW}_{1\ell m}\;
=0\cr}\;,\eqno(3.25)$$
\noindent
and
$$\partial _{t}\biggl[e^{(a-b)/2}\; h^{(-)RW}_{0\ell m}\biggr]\;
-\partial_{r}\biggl[e^{(b-a)/2}\; h^{(-)RW}_{1\ell m}\biggr]\; 
=\; 0\; .\eqno(3.26)$$
\noindent
Here,
$$\eqalign{F_{1\ell}(t,r)\;
=\;{{e^{a}}\over{r^{2}}}
&\biggl[{{4m}\over r} +\; 4m'\; -\; \ell(\ell +1)\biggr]\cr
&+\;e^{a-b}\biggl[{\ddot a} 
+{{1}\over{2}}\dot a (\dot a -\dot b)\biggr]\; + Z\cr}\;,\eqno(3.27)$$
\noindent
and
$$F_{2\ell}(t,r) \;
=\;-{{e^{b}}\over{r^{2}}}(\ell +2)(\ell -1)
+Z\;e^{b-a}\; +\ddot a +{1\over 2}\dot a(\dot a-\dot b)\; ,\eqno(3.28)$$
\noindent
with
$$Z
=-{{2e^{a}}\over r}
\biggl[m''\; +\;{{2(m')^{2}}\over{r^{2}}}\; e^{a}\; 
+\;{{2mm'}\over{r}}\; e^{a}\biggr]\; .\eqno(3.29)$$
\noindent
As usual, we define the 'mass function' $m(t,r)$ by
$$e^{-a(t,r)} 
=\; 1\; -{{2m(t,r)}\over{r}}\; .\eqno(3.30)$$
\noindent
The Einstein field equations imply
$$m' \; =\; 4\pi\; r^{2}\rho{\quad}.\eqno(3.31)$$
\noindent
Here, $\rho$ is the total energy density of all the radiative fields;
in the present case $\rho$ has contributions from $s=0$
(massless scalar), $s=2$ (graviton) and, if the Lagrangian 
contains Maxwell or Yang-Mills fields, also $s=1{\,}$.

The back-reaction due to the accumulated effects of the energy density
$\rho$ over very long time-scales, was discussed particularly in [17].
The net effect of the outgoing radiation on the background geometry 
$\gamma_{\mu\nu}$, which is spherically symmetric on averaging over 
time and angles, is to generate a space-time with (approximately) 
a Vaidya metric [18,19], depending on the mass function $m(t,r)$
above.  In this space-time, with metric written in the form of
Eq.(3.20), one has [17]:
$$e^{b(t,r)}\; 
=\;\biggl[{{\dot m}\over{f(m)}}\biggr]^{2}
\biggl[1\;-{{2m(t,r)}\over {r}}\biggr],\eqno(3.32)$$
\noindent
where $f(m)$ is, as yet, arbitrary, and where $m(t,r)$ obeys
$$m'\; =\;f(m)\biggl[1\;-{{2m(t,r)}\over{r}}\biggr]^{-1}.\eqno(3.33)$$
\noindent
The most natural choice for $f(m)$, which respects asymptotic flatness
and has purely outgoing radiation at large radii, is to take
$$f(m)\; =\; -\dot m{\quad}.\eqno(3.34)$$
This may be interpreted as the 'luminosity of the black hole'.  
A stellar-mass non-rotating black hole which only emits massless
particles loses mass at a rate $\dot m\propto -{\,}m^{-2}$ [20].
Following Eq.(3.30), a corresponding mass function which changes only
slowly with time must also change slowly with radius.  At a very late
time $t{\,}$, one expects that the mass function $m(t,r)$ is extremely
close to zero near the 'centre of symmetry' $r=0$ on the final
space-like hypersurface $\Sigma_F{\,}$, that $m(t,r)$ increases very
slowly with $r{\,}$, as one moves outward through the region containing
radiation, and that $m(t,r)$ settles at the total ADM 
(Arnowitt-Deser-Misner) mass [16] $M_I$ as $r\rightarrow\infty{\,}$.

As with the $s=0$ (scalar) perturbations, as treated in [2,3,21], 
we envisage making use of a Fourier-type expansion
(with respect to time) for the $s=2$ gravitational perturbations.  
For frequencies and angular momenta much stronger than those 
associated with the background space-time [for the $s=0$ scalar
case, see [3]. especially Eqs.(2.18,19,22)] therein, one has
$$\eqalignno{{\mid}k{\mid}\;&{\gg}\;
{{1}\over{2}}\;{\mid}{\dot a}-{\dot b}{\mid}{\quad},&(3.35)\cr
\ell(\ell +1)\; &{\gg}\; 2\, m'{\quad},&(3.36)\cr
m'\;&{\ll}\; 1{\quad}.&(3.37)\cr}$$
\noindent
In this case, we can again apply an adiabatic approximation for the
metric perturbations, in which all time derivatives of the
(background) metric components in the field equations are neglected;
correspondingly, by Eq.(3.34), all radial derivatives of the mass
function $m(t,r)$ are neglected, in leading order.  We apply this to
the odd-parity linearised Einstein equations (3.21-28).  (A similar
procedure will be adopted for the even-parity equations in Secs.7,8.)
The main consequence is that, in the well-known perturbation equations
on a Schwarzschild background [7], the Schwarzschild mass $M$ is
replaced by $m(r)$, namely, the mass inside a radius $r$. Since the
final space-like hypersurface $\Sigma_F$ is taken to be at an
extremely late time $T{\,}$, the black-hole radiation will only be 
present significantly on $\Sigma_F$ at very large radii $r{\,}$,
whence ${\,}a' + b'\;\simeq\; 0{\,}$ and
${\,}e^{b}\; \simeq \; e^{-a}{\,}$.  This is the regime studied 
principally in this paper, in order to compute the linearised classical
gravitational-wave solution subject to boundary data specified on
$\Sigma_I$ and $\Sigma_F$ (with $T$ rotated into the complex), and
hence the quantum amplitude for the final wave data.  The underlying 
approach is the same as in [2,3,17,21] for massless-scalar $(s=0)$ 
modes, and in [1] for $s=1$ Maxwell modes.
\end{section}

\begin{section}{Classical action for odd-parity metric perturbations}
We can now compute the odd-parity contribution to the classical
gravitational action [3].  In an arbitrary gauge, one has
$$S^{(2)}_{\rm class}\Bigl[h^{(1)}_{ij}\Bigr]\; =\; {1 \over{32\pi}}
\int_{\Sigma_F}\;d^{3}x\;\pi^{(1)ij}\;h^{(1)}_{ij}.\eqno(4.1)$$
\noindent
On discarding a total divergence, the spin-2 classical action can also
be written as
$$\eqalign{S^{(2)}_{\rm class}\;&=\; {1 \over
64\pi}\;\int_{\Sigma_F}\;d^{3}x\;
\sqrt{^{(3)}\gamma}\;n^{(0)\mu}\;\biggl(\bar h^{(1)\mu\nu}\; \nabla_{\alpha}
h^{(1)}_{\mu\nu}- 2h^{(1)}_{\alpha\nu}\;\nabla_{\rho}
h^{(1)\nu\rho}\biggr)\cr
&+{1 \over {16\pi}}\int_{\Sigma_F}\;d^{3}x \;\sqrt{^{(3)}\gamma}\;\;
{{N^{(1)}_{~~i}} \over{N^{(0)}}}\;\bar h^{(1)ik}_{~~~~~\mid k}.\cr}
\eqno(4.2)$$
Since odd- and even-parity perturbations are orthogonal, there are no
cross-terms in the action.  In an arbitrary gauge, the odd-parity
contribution to the classical gravitational action can be
re-written as 
$$\eqalign{S^{(2)}_{\rm class}\Bigl[(h^{(-)}_{ij})_{\ell m}\Bigr]
&={1\over{32\pi}} \int_{\Sigma_F}\; d^{3} x \;\sum_{\ell
\ell'mm'}\biggl(\pi^{(-)ij}\biggr)_{\ell m}\;
\biggl(h^{(-)}_{ij}\biggr)^{*}_{\ell' m'}\cr
& ={1 \over {32\pi}}\sum_{\ell m} \ell (\ell
+1)\;\int^{R\infty}_{0}\;dr\;h^{(-)*} _{1\ell
m}\;\biggl(\partial_{t}h^{(-)}_{1\ell m}+{2 \over r}h^{(-)}_{0\ell m}
-\partial_{r}h^{(-)}_{0\ell m}\biggr)\Bigl\arrowvert_{T}\cr
&+{1 \over 128 \pi}\sum_{\ell m}{{(\ell +2)!}\over {(\ell -2)!}}\;
\int^{R\infty}_{0}\;dr\;e^{a}{\;}{{h^{(-)*}_{2\ell
m}}\over{r^{2}}}\;\biggl(\partial_{t} h^{(-)}_{2\ell m}+2h^{(-)}_{0\ell
m}\biggr)\Bigl\arrowvert_{T},\cr}\eqno(4.3)$$
\noindent
Here, we have used Eqs.(3.7-9,15,16), and have taken the
perturbations to vanish initially.  Note that, if we were 
to evaluate Eq.(4.3) in the RW gauge, for which 
$h^{(-)RW}_{2\ell m} = 0{\,}$, so that the second integral 
would vanish, and then substitute for the Regge-Wheeler
functions {\it via} Eqs.(3.19-22), then, in the adiabatic 
aproximation, we would arrive back at Eq.(4.3) 
up to a boundary term
$$h^{(-)*}_{2\ell m}\;P^{(-)}_{\ell m}\Bigl\arrowvert
^{r=R_{\infty}}_{r=0}\eqno(4.4)$$
\noindent
where
$$P^{(-)}_{\ell m} \; =\; \ell (\ell
+1)\Biggl[\partial_{t}\;h^{(-)}_{1\ell m}
-r^{2}\;\partial_{r}\biggl({{h^{(-)}_{0\ell m}}\over
{r^{2}}}\biggr)\Biggr].\eqno(4.5)$$ 
\noindent
Thus, $S^{(2)}_{\rm class}[(h^{(-)}_{ij})_{\ell m}]$ is
gauge-invariant up to a boundary term.  We shall return below to the
question of boundary conditions for the odd-parity perturbations. 

At first sight, the odd-parity action looks unwieldy.  Ideally, we
would like to work with a classical action (both for odd and for even
parity) in the form $\int dr\;\psi(\partial_{t}\psi){\,}$, of the same
general kind as in the massless-scalar classical action in [2,3]. 
In the present gravitational case, $\psi$ would ideally also be
gauge-invariant and would obey a wave equation with a real potential.
To achieve this form, first use Eqs.(3.19-22) for the RW functions, and
substitute them into the field equation (3.25), to obtain
$$\eqalignno{\partial_{t}P^{(-)}_{\ell m}&=\ell (\ell + 1) \; F_{2
\ell}(r)\biggl[h^{(-)}_{1\ell m} +{1\over 2}\;
\partial_{r}\biggl({{h^{(-)}_{2\ell
m}}\over{r^{2}}}\biggr)\biggr],&(4.6)\cr
\partial_{r}P^{(-)}_{\ell m}&= -{{2\;P^{(-)}_{\ell m}}\over r}+ \ell
(\ell +1)\biggl({1\over2}F_{1\ell}(r)+{1\over
{r^{2}}}\biggr)\biggl(\partial_{t}h^{(-)}_{2\ell m}+2h^{(-)}_{0\ell
m}\biggr)\;.&(4.7)\cr}$$
\noindent
When we substitute into Eq.(4.3), using Eq.(4.6) for 
$h^{(-)}_{1\ell m}{\,}$, and then use Eqs.(3.28,4.7), 
the boundary term (4.4) vanishes.  As a consequence, 
we find in the adiabatic approximation that
$$S^{(2)}_{\rm class} \biggl[\Bigl(h^{(-)}_{ij}\Bigr)_{\ell m}\biggr]
 = -{1 \over
{32\pi}}\sum^{\infty}_{\ell =2}\;\sum^{\ell}_{m=-\ell}{{(\ell
-2)!}\over{(\ell +2)!}} \;
\int^{R_{\infty}}_{0}\;dr\;e^{a}\xi^{(-)}_{2\ell
m}\Bigl(\partial_{t}\xi^{(-)*}_{2\ell m}\Bigr)\Bigl\arrowvert_{
t=T},\eqno(4.8)$$
\noindent
where $\xi^{(-)}_{2\ell m}$ is defined by
$$\xi^{(-)}_{2\ell m} =  r\;P^{(-)}_{\ell m}.\eqno (4.9)$$
\indent
Eqs.(3.19-21) show that $\xi^{(-)}_{2\ell m}$ is gauge-invariant.
Indeed, $\xi^{(-)}_{2\ell m}$ is related to Moncrief's [15]
gauge-invariant generalisation of the Zerilli function, 
$Q^{(-)}_{\ell m}$ [22,23], defined as
$$Q^{(-)}_{\ell m} ={{e^{-a}}\over r}\Biggl[h^{(-)}_{1\ell m} +{1
\over 2}\; r^{2}\;\partial_{r}\biggl({{h^{(-)}_{2\ell m}}\over
{r^{2}}}\biggr)\Biggr] , \eqno(4.10)$$
\noindent
by
$$Q^{(-)}_{\ell m} =-{{(\ell -2)!}\over {(\ell +2)!}}\;\Bigl(\partial
_{t}\xi^{(-)}_{2\ell m}\Bigr)\;,\eqno(4.11)$$
\noindent
which, in effect, replicates Eq.(4.6).  Note the simplifying property
of $Q^{(-)}_{\ell m}$, namely, that it is written entirely in terms of
perturbations of the 3-geometry (our chosen boundary data).  Further,
$Q^{(-)}_{\ell m}$ is automatically gauge-invariant, since it is
independent of the perturbed lapse and shift.

In the adiabatic approximation, the function $\xi^{(-)}_{\ell m}$
obeys the wave equation, of RW type:
$$e^{-a}\;\partial_{r}\biggl[e^{-a}\;\Bigl(\partial_{r}\xi^{(-)}_{2\ell
m}\Bigr)\biggr]-(\partial_{t})^{2}\xi^{(-)}_{2\ell m} -V^{(-)}_{\ell}
(r)\;\xi^{(-)}_{\ell m}\;=\;0,\eqno(4.12)$$
\noindent
where
$$V^{(-)}_{\ell}(r) 
=\;e^{-a}\biggl[{{\ell(\ell +1)}\over{r^{2}}}
-{{6\;m(r)}\over{r^{3}}}\biggr]{\quad}>{\;}{\,}0{\quad}.\eqno(4.13)$$
\noindent
Further, in the adiabatic approximation, the function 
$Q^{(-)}_{\ell m}$ obeys the same equation (4.12). In the RW gauge, 
one would solve for $Q^{(-)RW}_{\ell m}{\,}$, then determine 
$h^{(-)RW}_{1\ell m}$ from Eq.(4.10), and then determine 
$h^{(-)RW}_{0\ell m}$ with the help of Eq.(3.26).
\end{section}

\begin{section}{ Boundary conditions for odd-parity perturbations}
In classical Lorentzian general relativity, one would expect to choose
regular Cauchy data on an initial space-like hypersurface 
$\Sigma_I{\,}$, which would then evolve smoothly into 
$\{x^{0}{\,}>{\,}0\}$, subject to the linear hyperbolic equation
(4.12).  A natural initial condition, for given quantum numbers 
$\ell{\,} m$ [24], would be to assume an initially stationary 
odd-parity multipole:
$$\partial_{t}\xi^{(-)}_{2\ell m}\;\mid_{t=0}\; = 0.\eqno(5.1)$$

The combined Einstein/massless-scalar boundary-value problem,
originally posed in [2,3], for complex time-separation 
$T={\mid}T{\mid}\exp(-i\theta){\;},{\;}0<\theta\leq\pi/2{\,}$,
involved specifying the intrinsic 3-metric $(h_{ij})_{I,F}$
and the value of the scalar field $(\phi)_{I,F}$ on the initial and
final space-like hypersurfaces $\Sigma_{I}{\,},\Sigma_F{\,}$.  
By Eq.(4.11), the above Eq.(5.1) reads
$$Q^{(-)}_{\ell m}(0,r)\; =\; 0\eqno(5.2)$$
\noindent
or, equivalently,
$$\eqalignno{h^{(-)}_{1\ell m}(0,r)\; &=\; 0,&(5.3)\cr
h^{(-)}_{2\ell m}(0,r)\; &=\; 0,&(5.4)\cr}$$
\noindent
We therefore take these as our (Dirichlet) boundary conditions on the
odd-parity gravitational perturbations, on the initial surface
$\Sigma_I{\,}$, even though they may have arisen from consideration 
of the Cauchy problem.

In [2,3,21] for the $s=0$ case and in [1] for the $s=1$ (Maxwell)
case, we made use of the adiabatic approximation in order to
separate the perturbation problem with respect to the variables $t$
and $r$.  Here, for $s=2{\,}$, we first separate the odd-parity 
Eqs.(3.24,25) in the RW gauge, and then use Eqs.(3.19-22) 
to determine the time-dependence (in particular) in any gauge.

As in the massless-scalar $(s =0)$ case, we introduce a 'Fourier-type
expansion' 
$$h^{(-)RW}_{0,1,2\ell m}(t,r) =\;\int^{\infty}_{-\infty}\;
dk\;a^{(-)}_{k\ell m}\; h^{(-)RW}_{0,1,2k\ell m}(t,r),\eqno(5.5)$$
\noindent
where the $\{a^{(-)}_{k\ell m }\}$ are certain odd-parity 'Fourier'
coefficients.  With suitable treatment of any arbitrary phase factors
involved, in order to separate the odd-parity field equations (3.21,22)
in the adiabatic approximation, one must have 
$$\eqalignno{h^{(-)RW}_{0\ell m} (t,r)&\propto\;\cos (kt),&(5.6)\cr
h^{(-)RW}_{1\ell m} (t,r)&\propto\;\sin (kt),&(5.7)\cr}$$
\noindent
(Of course, the normal-mode $e^{-ikt}$ time dependence for the
functions $h^{(-)RW}_{0,1,\ell m}$ also satisfies the field
equations.)  In Eq.(3.22), if ${\,}h^{(-)RW}_{0\ell m}{\,}$, which is 
related to the odd-parity shift and can thus be freely specified, 
has $\cos (kt)$ time dependence, then $h^{(-)}_{0\ell m}$ must have 
the same time-dependence, while $\Lambda_{\ell m}(t,r)$ must have 
$\sin(kt)$ time dependence.  But, by Eq.(3.19), $h^{(-)}_{2\ell m}$ 
must then have $\sin(kt)$ time-dependence.  Similarly, from Eq.(3.20), 
given that $h^{(-)RW}_{1\ell m}$ has $\sin(kt)$ time-dependence, 
$h^{(-)}_{1\ell m}$ must also have $\sin(kt)$ time-dependence, 
as must $\Lambda_{\ell m}(t,r){\,}$.  These conclusions are 
consistent with our choice of boundary conditions (5.3,4).  
Noting Eqs.(4.5,10,11), the Dirichlet conditions (5.3,4) are
equivalent to the boundary condition (5.1), which is analogous to a
specification of momenta in a 
$(\xi^{(-)}_{2\ell m},\;\partial_{t}\xi^{(-)}_{2\ell m})$ 
representation.  This also accounts for the minus sign in Eq.(4.8).
\end{section}

\begin{section}{Asymptotically-flat gauge}
For large $r{\,}$, the potential term in Eq.(4.12) vanishes 
sufficiently rapidly that $\xi^{(-)}_{2 \ell m}$ becomes a 
superposition of outgoing and ingoing waves at radial infinity.  
Note that $Q^{(-)}_{\ell m}$ also obeys Eq.(4.12); thus, Eq.(4.10) 
in the RW gauge tells us that 
$h^{(-)RW}_{1\ell m}= rQ^{(-)RW}_{\ell m}e^{a}= O(r)$ at large $r$.  
Now, the field equation (3.26) implies that 
$$\partial_{t}h^{(-)RW}_{o\ell m} \; =\; e^{-a}\;\partial_{r}\biggl(r
\;Q^{(-)RW}_{\ell m}\biggr).\eqno(6.1)$$
\noindent
That is, odd-parity metric perturbations diverge at large $r{\,}$, 
in the RW gauge.  This is only a coordinate effect, as the 
Riemann-curvature invariants decay at a rate $O(r^{-1})$ at large 
$r$ [8,9].  (A similar phenomenon occurs for the even-parity 
perturbations in the RW gauge.)  Here, in the odd-parity case, 
we construct a gauge transformation to an asymptotically-flat (AF) 
gauge, in which the radiative behaviour of the metric perturbations 
becomes manifest.

Our odd-parity AF gauge is chosen such that
$$h^{(-)AF}_{0\ell m}(t,r) \; =\; 0.\eqno(6.2)$$
\noindent
Thus, in terms of the preceding RW gauge:
$$\eqalignno{h^{(-)AF}_{0\ell m}\;& = \; 0\; =\; h^{(-)RW}_{0\ell m}\;
-\partial_{t}\Lambda_{\ell m},&(6.3)\cr
h^{(-)AF}_{1\ell m}\; &= h^{(-)RW}_{1\ell
m}-\partial_{r}\Lambda_{\ell m}+\; {{2\;\Lambda_{\ell m}}\over
r},&(6.4)\cr
h^{(-)AF}_{2\ell m}\; &= \; 2\Lambda_{\ell m}.&(6.5)\cr}$$
\noindent
Given $h^{(-)RW}_{0\ell m}$ and $h^{(-)RW}_{1\ell m}$  as a
starting-point, one can, from the above, determine $\Lambda_{\ell
m}(t,r)$ and hence $h^{(-)AF}_{1 \ell m}$ and $h^{(-)AF}_{2\ell m}$.
On substituting for $h^{(-)RW}_{1 \ell m}$ from Eq.(6.4) into Eq.(3.22),
one finds
$$(\partial_{t})^{2}h^{(-)AF}_{1\ell m}\; =\; -{{2\;\lambda \;
e^{-a}}\over {r^{2}}}\; h^{(-)RW}_{1 \ell m}.\eqno(6.6)$$
\noindent
Now, following the approach used throughout when studying boundary
conditions at the final surface $\Sigma_{F}\;(t=T)$, set:
$$h^{(-)AF}_{1\ell m}(t,r)\; =\; \int^{\infty}_{-\infty}\;
dk\;a^{(-)}_{k\ell m}\;h^{(o)AF}_{1k\ell}(r)\;{{\sin(kt)}\over
{\sin(kT)}},\eqno(6.7)$$
\noindent
where the $\{h^{(-)AF}_{1k\ell}(r)\}$ are real radial functions, and
where ${\,}a^{(-)*}_{k\ell m} =(-1)^{m}\; a^{(-)}_{-k\ell ,-m}{\,}$.
Similarly, construct a corresponding expansion for 
$Q^{(-)RW}_{\ell m}(t,r)$.  Then, from Eq.(6.6), one has
$$h^{(-)AF}_{1\ell m}(t,r) \; =\; {{2\lambda}\over r}\;
\int^{\infty}_{-\infty} \;dk\;{{a^{(-)}_{k\ell m}}\over{k^{2}}}\;
Q^{(-)RW}_{k\ell}(r)\; {{\sin (kt)}\over {\sin(kT)}}\;,\eqno(6.8)$$
\noindent
which is $O(r^{-1})$ at large $r{\,}$, as required.  
On using Eq.(4.5), one further finds
$$\eqalign{\xi^{(-)AF}_{2 \ell m}(t,r) \;& =\; r\;\ell\;(\ell
+1)\;\Bigl(\partial_{t}h^{(-)AF}_{1\ell m}\Bigr)\cr
&=\int^{\infty}_{-\infty}\;dk\;\hat
a^{(-)}_{2k\ell m}\;\xi^{(-)AF}_{2k\ell}(r)\; {{\cos(kt)}\over
{\sin(kT)}},\cr}\eqno(6.9)$$
\noindent
where
$$\hat a^{(-)}_{2k\ell m} \; =\; k\; \ell \;(\ell +1)\; a^{(-)}_{k\ell
m},\eqno(6.10)$$ 
\noindent 
and where
$$\xi^{(-)AF}_{2k\ell}(r)\; =\; r\; h^{(-)AF}_{1k\ell}(r)\eqno(6.11)$$
\noindent
satisfies
$$e^{-a}\;\biggl(e^{-a}\;\xi^{(-)AF'}_{2k\ell}\biggr)'\; +\; \biggl[k^{2}
-V^{(-)}_{\ell}(r)\biggr] \xi^{(-)AF}_{2k\ell} = 0.\eqno(6.12)$$

As in the spin-0 case [3] and in the spin-1 case [1], we have, 
for ${\,}k>0{\,}$:
$$\eqalignno{\xi^{AF}_{2k\ell-}(r)\; &\sim\;r\; j_{\ell}(kr),\;
r\rightarrow 0 &(6.13)\cr
\xi^{AF}_{2k\ell-}(r)\; &\sim\;\biggl(z_{2k\ell-}\exp
\Bigl(ikr^{*}_{s}\Bigr)\; +
z^{*}_{2k\ell-}\;\exp\Bigl(-ikr^{*}_{s}\Bigr)\biggr),\;\;r^{*}_{s}
 \rightarrow \infty,&(6.14)\cr}$$
\noindent
where the $j_{\ell}(z)$ are spherical Bessel functions, and 
where $r^{*}_{s}$ is the Schwarzschild Regge-Wheeler coordinate [7,16]:
$$r^{*}_{s}\; =\; r +\; 2M\;\ln\biggl[(r-2M)/2M\biggr].\eqno(6.15)$$
\noindent
Here, $M$ denotes the ADM (Arnowitt-Deser-Misner) mass of the
'space-time', as measured at spatial infinity [16].
Thence, one deduces the normalisation property
$$\int^{R\infty}_{0}\;dr\;e^{a}\;\xi^{AF}_{2k\ell -}(r)\;
\xi^{AF}_{2k'\ell-}(r)\Bigl\arrowvert _{\Sigma_F}\; =\;2\;\pi\;
{\mid} z_{2k\ell -}{\mid}^{2}\;
\biggl[\delta(k,k')+\delta(k,-k')\biggr]\;.\eqno(6.16)$$
\noindent
The resulting form of the classical action for odd-parity (spin-2)
gravitational perturbations can then be expressed as a functional of
the complex quantities $\{a_{2k\ell m-}\}$ which encode the boundary
data on $\Sigma_F$ for the odd-parity gravitational
perturbations. Here,
$$\eqalign{&S^{(2)}_{\rm class}\biggl[\{a_{2k\ell m-}\}\biggr]\cr
 &={1 \over
16}\;\sum^{\infty}_{\ell =2}\; \sum^{\ell}_{m=-\ell}\;{{(\ell
-2)!}\over {(\ell +2)!}}\; \int^{\infty}_{0}\;dk{\;}{\,}k\;
{\mid}z_{2k\ell-}{\mid}^{2}{\quad}{\mid}(a_{2k\ell m-}){\,}
-(a_{2,-k\ell m-}){\mid}^{2}\;\cot(kT){\quad}.\cr}\eqno(6.17)$$
\noindent
From this expression, one proceeds as in [3] (for spin-0) and [1] 
(for spin-1) to study the semi-classical quantum amplitude or wave
function, proportional to $\exp(iS^{(2)}_{\rm class}),$ as a function
of the complexified time-interval $T={\mid}T{\mid}\exp(-i\theta)$,
for ${\,}0<\theta\leq\pi/2{\,}$.  Just as in the spin-0 and spin-1 case,
one straightforwardly recovers the complex Lorentzian amplitude for
odd-parity gravitational perturbations, on taking the limit 
$\theta\rightarrow 0_{+}{\;}$.
\end{section}

\begin{section}{Even-parity perturbations}
Working with even-parity gravitational perturbations in the RW
formalism is notoriously more difficult than working with those 
of odd parity.  Yet, Chandrasekhar [25] showed that solutions to 
Zerilli's even-parity equation [22,23] [Eq.(8.10) below] can be 
expressed in terms of the odd-parity solutions.  One might therefore 
expect that our results for the even-parity action should mirror 
those for the odd-parity action.

We expand the even-parity perturbations as
$$\eqalign{\Bigl(h^{(+)}_{ij}\Bigr)_{\ell m}(x)\;&=\;h^{(+)}_{1\ell
  m}(t,r)\biggl[(f_{1})_{ij}\biggr]_{\ell m}\;+\;H_{2\ell m}(t,r)\;
  e^{(a-b)/2}\;\biggl[(f_{2})_{ij}\biggr]_{\ell m}\cr
&+\; r^{2}\; K_{\ell m}(t,r)\;\biggl[(f_{3})_{ij}\biggr]_{\ell m}\; +\;
  r^{2}G_{\ell m}(t,r)\;\biggl[(f_{4})_{ij}\biggr]_{\ell
  m}\;,\cr}\eqno(7.1)$$
\noindent
Here, the non-zero components of the (un-normalised) basis of the
symmetric tensor spherical harmonics 
$((f_{1,2,3,4})_{ij})_{\ell m}$ are defined by 
$$((f_1)_{r\theta})_{\ell m}{\;}
={\;}({\partial}_{\theta}Y_{\ell m}){\quad},\eqno(7.2)$$
$$((f_1)_{r\phi})_{\ell m}{\;}
={\;}({\partial}_{\phi}Y_{\ell m}){\quad},\eqno(7.3)$$
$$((f_2)_{rr})_{\ell m}{\;}
={\;}Y_{\ell m}{\quad},\eqno(7.4)$$
$$((f_3)_{\theta\theta})_{\ell m}{\;}
={\;}Y_{\ell m}{\quad},\eqno(7.5)$$
$$((f_3)_{\phi\phi})_{\ell m}{\;}
={\;}({\sin}^{2}\theta){\,}Y_{\ell m}{\quad},\eqno(7.6)$$
$$((f_4)_{\theta\theta})_{\ell m}{\;}
={\;}({\partial}_{\theta})^{2}Y_{\ell m}{\quad},\eqno(7.7)$$
$$((f_4)_{\theta\phi})_{\ell m}{\;}
={\;}\bigl({\partial}_{\theta}{\partial}_{\phi}{\,}
-{\,}(\cot\theta){\partial}_{\phi}\bigr)Y_{\ell m}{\quad},\eqno(7.8)$$
$$((f_4)_{\phi\phi})_{\ell m}{\;}
={\;}\bigl(({\partial}_{\phi})^{2}{\,}
+{\,}(\sin\theta{\,}\cos\theta){\,}{\partial}_{\theta}\bigr)Y_{\ell m}
{\quad},\eqno(7.9)$$
 
The non-zero inner products are given by
$$\eqalignno{\int\;d\Omega\;\Bigl[(f_{1})^{ij}\Bigr]_{\ell
    m}\;\Bigl[(f_{1})_{ij}\Bigr]^{*}_{\ell'm'} \;&=\;{{2\; e^{-a}}\over
    {r^{2}}}\; \ell(\ell +1)\;\delta_{\ell
    \ell'}\;\delta_{mm'}\;,&(7.10)\cr
\int\;d\Omega\;\Bigl[(f_{2})^{ij}\Bigr]_{\ell
    m}\;\Bigl[(f_{2})_{ij}\Bigr]^{*}_{\ell'm'} \;&=\;e^{-2a}\;
\delta_{\ell
    \ell'}\;\delta_{mm'}\;,&(7.11)\cr
\int\;d\Omega\;\Bigl[(f_{3})^{ij}\Bigr]_{\ell
    m}\;\Bigl[(f_{3})_{ij}\Bigr]^{*}_{\ell'm'} \;&=\; {2 \over{r^{4}}}
\delta_{\ell
    \ell'}\;\delta_{mm'}\;,&(7.12)\cr
\int\;d\Omega\;\Bigl[(f_{4})^{ij}\Bigr]_{\ell
    m}\;\Bigl[(f_{4})_{ij}\Bigr]^{*}_{\ell'm'}
    \;&=\;\Bigl({\Lambda_{\ell}(\Lambda_{\ell}-1)\over{r^{4}}}\Bigr) 
\delta_{\ell
    \ell'}\;\delta_{mm'}\;,&(7.13)\cr
\int\;d\Omega\;\Bigl[(f_{3})_{ij}\Bigr]_{\ell
    m}\;[(f_{4})^{ij}]^{*}_{\ell'm'}
    \;&=\;-{{\ell(\ell+1)}\over{r^{4}}} \;
\delta_{\ell
    \ell'}\;\delta_{mm'}\;,&(7.14)\cr}$$
\noindent
where we define $\Lambda_{\ell} = \ell (\ell +1)$.
The even-parity basis is also orthogonal to the odd-parity 
basis of Sec.3.

Further, for the even-parity perturbed shift, one can write
$$\Bigl[N^{(+)}_{i}\Bigr]_{\ell m}\; =\; \biggl[H_{1\ell m}(t,r)\;
Y_{\ell m}\;,\;\;
h^{(+)}_{0\ell m}(t,r)\;(\partial_{\theta}Y_{\ell m}),
h^{(+)}_{0\ell m}(t,r)\;(\partial_{\phi}Y_{\ell m})\biggr].\eqno(7.15)$$
\noindent
For the perturbed lapse,
$$\Bigl[N^{(1)(+)}\Bigr]_{\ell m}\; =\; -{1 \over 2}\;H_{0\ell
    m}(t,r)\;e^{-{1 
    \over 2}a}\; Y_{\ell m}.\eqno(7.16)$$
\noindent
Again, $H^{*}_{0\ell m}{\,}=(-1)^{m}\;H_{0\ell,-m}$, etc.  Hence, for
the linear-order perturbation $h^{(1)}_{\mu\nu}$ of the 4-dimensional
metric, the quantities $h^{(1)}_{tt},h^{(1)}_{rr}$ and $h^{(1)}_{tr}$
behave as scalars under rotations (their odd-parity part vanishes),
while $h^{(1)}_{t\theta}, h^{(1)}_{t\phi},h^{(1)}_{r\theta},$ and 
$h^{(1)}_{r\phi}$ behave as vectors and, for $a\;
=\;\theta,\phi,\;b\;=\theta,\phi,$ the $2\times 2$ angular block
$h^{(1)}_{ab}$ is a tensor under rotations.

The even-parity gravitational momentum components can,
correspondingly, be written in the form
$$\eqalign{(\pi^{(+)}_{ij})_{\ell m}\; &=\; (^{3}\gamma)^{1 \over 2}\;
  \biggl(P_{h 1\ell m}(t,r)\;\Bigl[(f_{1})_{ij}\Bigr]_{\ell m}\;
  +\;P_{H2\ell m}(t,r)\;\Bigl[(f_{2})_{ij}\Bigr]_{\ell m}\cr
&+r^{2}\;P_{K\ell m}(t,r)\;\Bigl[(f_{3})_{ij}\Bigr]_{\ell
  m}\;+\;r^{2}\;P_{G\ell m}(t,r)\;\Bigl[(f_{4})_{ij}\Bigr]_{\ell
  m}\biggr).\cr}\eqno(7.17)$$
\noindent
Again, one can easily show that the $P$'s in Eq.(7.17) are related to
$h_{1},H_{2}, K,$ and $G$ of the corresponding Eq.(7.1)
$$\eqalignno{P_{h1\ell m}(t,r)\;&=\;{1\over2}\;e^{a/2}
      \biggl[\bigl(\partial_{t}h^{(+)}_{1\ell
      m}\bigr)\;-r^{2}\;\partial _{r}\;\biggl({{h^{(+)}_{0\ell
      m}}\over {r^{2}}}\biggr)\biggr],&(7.18)\cr
P_{G\ell m}(t,r)\;&=\;{1\over2}\;e^{a/2}
      \biggl((\partial_{t}G_{\ell
      m})\;-{{2h^{(+)}_{0\ell
      m}}\over {r^{2}}}\biggr),&(7.19)\cr
P_{K\ell m}(t,r)\;&=\;-{1\over2}\;e^{a/2}
      \Biggl[(\partial_{t}H_{2\ell
      m}\bigr)\;+\;(\partial_{t}K_{\ell m})\;+\;\biggl(a' -{2 \over
      r}\biggr) \;e^{-a}\;H_{1\ell m}\cr
&-2\;e^{-a}\;\bigl(\partial_{r}H_{1\ell m}\bigr)\;+\;{{2\;\ell (\ell
      +1)\;h^{(+)}_{0\ell m}}\over
      {r^{2}}}\;-\;\ell(\ell+1)\;\bigl(\partial_{t}G_{\ell
      m}\bigr)\Biggr],&(7.20)\cr
P_{H2\ell m}(t,r)\;&=\;-e^{a/2}\;\bigl(\partial_{t}K_{\ell
      m}\bigr)\;+\;{2 \over r}\;e^{-a/2}\;H_{1\ell
      m}\;-\;{{\ell(\ell+1)\;h^{(+)}_{0\ell
      m}}\over{r^{2}}}\;e^{a/2}\cr
&+{1 \over2}\ell(\ell+1)\;e^{a/2}
      \bigl(\partial_{t}G_{\ell
      m}\bigr).&(7.21)\cr}$$

For even-parity gravitational perturbations, gauge transformations are
induced by even-parity gauge vector fields $(\xi^{(+)\mu})_{\ell m}$,
of the form:
$$\eqalign{\Bigl(\xi^{(+)t}\Bigr)_{\ell m}\;&=\;X^{(+)}_{0\ell
    m}(t,r)\;Y_{\ell 
    m}\;,\quad\Bigl(\xi^{(+)r}\Bigr)_{\ell m}\;=\;X^{(+)}_{1\ell m}
(t,r)\;Y_{\ell
    m}\;,\cr
\Bigl(\xi^{(+)\theta}\Bigr)_{\ell m}\;&=\;X^{(+)}_{2\ell
    m}(t,r)\;\partial_{\theta} Y_{\ell
    m}\;,\quad\Bigl(\xi^{(+)\phi}\Bigr)_{\ell m}\;\;=\;{{X^{(+)}_{2\ell
    m}(t,r)}\over{\sin^{2}\theta}} (\partial_{\phi}Y_{\ell
    m}).\cr}\eqno(7.22)$$
\noindent
Within the adiabatic approximation, these induce the following
even-parity gauge transformations:
$$\eqalignno{H'_{0\ell m}\;&=\;H_{0\ell m}\;-\;a'X^{(+)}_{1\ell
    m}\;+\;2\Bigl(\partial_{t}X^{(+)}_{0\ell m}\Bigr),&(7.23)\cr
H'_{1\ell m}\;&=\;H_{1\ell m}\;-\;e^{-a}\;\Bigl(\partial_{r}X^{(+)}_{0\ell
    m}\Bigr)\;-\;e^{a} \Bigl(\partial_{t}X^{(+)}_{1\ell m}\Bigr),&(7.24)\cr
H'_{2\ell m}\;&=\;H_{2\ell m}\;-\;a'X^{(+)}_{1\ell
    m}\;-\;2\Bigl(\partial_{r}X^{(+)}_{1\ell m}\Bigr),&(7.25)\cr
K'_{\ell m}\;&=\; K_{\ell m}\;-\; {{2\;X^{(+)}_{1\ell m}}\over r},&(7.26)\cr
G'_{\ell m}\;&=\;G_{\ell m}\;-\;2X^{(+)}_{2\ell m},&(7.27)\cr
h^{(e)'}_{0\ell m}\;&=\;h^{(+)}_{0\ell m}\;+\;e^{-a}\;X^{(+)}_{0\ell
    m}\;-\; r^{2}\Bigl(\partial_{t}X^{(+)}_{2\ell m}\Bigr),&(7.28)\cr
h^{(+)'}_{1\ell m}\;&=\;h^{(+)}_{1\ell m}\;-\;e^{a}\;X^{(+)}_{1\ell
    m}\;-\; r^{2}\Bigl(\partial_{r}X^{(+)}_{2\ell m}\Bigr).&(7.29)\cr}$$

As in the odd-parity case, we would like to construct gauge-invariant
combinations of components of the perturbed 3-geometry.  Following 
[15], we define
$$\eqalignno{k_{1\ell m}\;&=\; K_{\ell
    m}\;+\;e^{-a}\;\Biggl(r(\partial_{r}G_{\ell
    m})\;-\;{{2\;h^{(+)}_{1\ell m}}\over r}\Biggr),&(7.26)\cr
k_{2\ell m}\;&=\;{1\over 2}\Biggl[e^{a}\;H_{2\ell m}\;
    -\;e^{a/2}\;\partial_{r}\biggl(r\;e^{a/2}\;K_{\ell
    m}\biggr)\Biggr]
&(7.27)\cr}$$
\noindent
It can be shown that both the functions $k_{1\ell m}$ and 
$k_{2\ell m}$ are gauge-invariant.  For future use, in the calculation 
of the even-parity classical action, we define [15] the linear 
combination of $k_{1\ell m}$ and $k_{2\ell m}$:
$$q_{1\ell m}\;=\; r\;\ell (\ell +1)\; k_{1\ell
    m}\;+\;4\;r\;e^{-2a}\;k_{2 \ell m}\;.\eqno(7.32)$$

At this stage, as with the odd-parity case, we again make use of the
property of the uniqueness of the (even-parity) RW gauge.  In the RW
gauge, one has
$$\eqalign{H^{RW}_{0\ell m}\;&=\;H_{0\ell m}\; -\;{1\over
    2}\;r^{2}\;a'\;e^{-a}\;\biggl({{2\;h^{(+)}_{1\ell
    m}}\over{r^{2}}} \;-\;(\partial_{r}G_{\ell m})\biggr)\;
    +\;r^{2}e^{a} \;(\partial_{t})^{2}G_{\ell m}\cr
&-2\;e^{a}(\partial_{t}h^{(+)}_{0\ell m}),\cr}\eqno(7.33)$$
\noindent
Then,
$$\eqalign{H^{RW}_{1\ell m}\;&=\;H_{1\ell
    m}\;+r^{2}(\partial_{r}\partial_{t}G_{\ell
    m})(\partial_{r}h^{(+)}_{0\ell m})\;-\;
(\partial_{t}h^{(+)}_{1\ell m})\;+\;r\;\biggl(1\;+\;{1 \over
    2}\;r\;a'\biggr) (\partial_{t}G_{\ell m})\cr
&-\;a'\;h^{(+)}_{0\ell m}.\cr}\eqno(7.34)$$

Next,
$$\eqalign{H^{RW}_{2\ell m}&=H_{2\ell m}\;+\; \biggl(a'\;-\;{4 \over
    r}\biggr)\;e^{-a}\;\biggl(h^{(+)}_{1\ell m}\; -{1 \over
    2}\;r^{2}\;(\partial_{r}G_{\ell m})\biggr)\cr
&+\;r^{2}\;e^{-a}\;\Biggl[(\partial_{r})^{2}G_{\ell
    m}\;-\;2{\,}\partial_{r}\biggl({{h^{(+)}_{1\ell m}}\over
    {r^{2}}}\biggr)\Biggr].\cr}\eqno(7.35)$$
\noindent
Further,
$$K^{RW}_{\ell m}\;=\;K_{\ell m}\;-\;{{2\;e^{-a}}\over
    r}\;\bigl(h^{(+)}_{1\ell m}\; -{1 \over
    2}\;r^{2}\;(\partial_{r}G_{\ell m})\bigr);\eqno(7.36)$$
\noindent
with
$$G^{RW}_{\ell m}\;=\;0\;=\;G_{\ell m}\;-2X^{(+)}_{2\ell m}\;,\eqno(7.37)$$
\noindent
together with
$$h^{(+)RW}_{0\ell m}\;=\;0,\eqno(7.38)$$
\noindent
and
$$h^{(+)RW}_{1\ell m}\;=\;0,\eqno(7.39)$$
\noindent
where (in the RW gauge)
$$\eqalignno{X^{(+)}_{0\ell m} &=\; e^{-b}\;\biggl({1 \over
    2}\;r^{2}\;(\partial_{t}G_{\ell m})\;-\;h^{(+)}_{0\ell
    m}\biggr),&(7.40)\cr
X^{(+)}_{1\ell m} &=\; e^{-a}\;\biggl(h^{(+)}_{1\ell
    m}\;-\;  {1 \over
    2}\;r^{2}\;(\partial_{r}G_{\ell m})\biggr),&(7.41)\cr
X^{(+)}_{2\ell m} &=\; {1\over 2}\;G_{\ell m}\;.&(7.42)\cr}$$

At late times, following the gravitational collapse to a black
hole, in the absence of background matter and in the adiabatic
approximation, the even-parity RW field equations are seven coupled
equations for the four unknowns 
$(H^{RW}_{0\ell m},H^{RW}_{1\ell m},H^{RW}_{2\ell m},K^{RW}_{\ell m})$.  
Assuming that ${\,}\ell{\,}\geq{\,}2{\;}$
--- that is, that we are studying dynamical modes --- we give here
those RW field equations which are of first order in $r$ and $t{\,}$ 
[23].  These are, respectively, the $(t\theta),(tr)$ and $(r\theta)$
components of the linearised field equations:
$$\eqalignno{\biggl(\partial_{r}H^{RW}_{1\ell m}\biggr)\;
  &+\;{{2m}\over{r^{2}}} \; e^{a}\;H^{RW}_{1\ell m}\;
  =\;e^{a}\partial_{t}\biggl(K^{RW}_{\ell m}\;+\;H^{RW}_{2\ell m}
\biggr),&(7.43)\cr
{1\over2}\;\ell(\ell +1)\;H^{RW}_{1\ell m}&=\;
  -r^{2}(\partial_{t}\partial_{r}\;K^{RW}_{\ell m})\;+\;r^{2}
\biggl(\partial_{t} 
H^{RW}_{0\ell m}\biggr)\cr
&-r\;e^{a}\;\biggl(1-{{3m}\over r}\biggr)\biggl(\partial_{t}K^{RW}_{\ell
  m}\biggr),&(7.44)\cr
\biggl(\partial_{t}H^{RW}_{1\ell m}\biggr)&=\;e^{-a}
\biggl(\partial_{r}H^{RW}_{0\ell
  m}\biggr)\;-\;e^{-a}\biggl(\partial_{r}K^{RW}_{\ell m}\biggr)
\;+\; {{2m}\over{r^{2}}}
H^{RW}_{0\ell m},&(7.45)\cr}$$
\noindent
and the $(\theta \phi)$ component
$$H^{RW}_{0\ell m}{\quad}={\quad}H^{RW}_{2\ell m}{\quad}
\equiv{\quad}H^{RW}_{\ell m}{\quad}.\eqno(7.46)$$
\noindent
We also give one second-order equation, namely, the $(rr)$ component:
$$\eqalign{e^{2a}\bigl(\partial{_t}\bigr)^{2}K^{RW}_{\ell m}\;
&=\;{2\over r}\;
  e^{a} 
\biggl(\partial_{t}H^{RW}_{1\ell m}\biggr)\;
-{1\over r}
\biggl(\partial_{r}H^{RW}_{2\ell m}\biggr)\;+\;{{e^{a}\over r}}\;
\biggl(1-{m\over
  r}\biggr)
\biggl(\partial_{r}K^{RW}_{\ell m}\biggr)\cr
&-{{e^{a}}\over{2\;r^{2}}}(\ell +2)(\ell -1)\;\biggl(K^{RW}_{\ell m}\;-\;
H^{RW}_{2\ell m}\biggr)\;.\cr}\eqno(7.47)$$

Following Eq.(7.44), we find, for the gauge-invariant component defined
in Eq.(7.32): 
$$(\partial_{t}q_{1\ell m})\; =\; \ell(\ell
+1)\;\biggl(r(\partial_{t}K^{RW}_{\ell m})\; -e^{-a}\; H^{RW}_{1\ell
  m}\biggr).\eqno(7.48)$$
\noindent
We also find
$$(\partial_{r}q_{1\ell m})\;=\; -\ell (\ell
+1)\;\Biggl[2\;e^{-a}\;k_{2\ell m}\;+\;\biggl(1\;+\;{1\over
    2}\;r\;a'\biggr)k_{1\ell m}\Biggr],\eqno(7.49)$$
\noindent
where, in the RW gauge,
$$q_{1\ell m}\;\equiv\;2\;r\;e^{-a}H^{RW}_{\ell
  m}\;-\;2\;r^{2}\;e^{-a}\biggl(\partial_{r}K^{RW}_{\ell m}\biggr)
\;+\;2(\lambda
  \;r\;+\;3 \;m)K^{RW}_{\ell m}\eqno(7.50)$$
\noindent
with ${\,}\lambda{\;}={\;}{{1}\over{2}}(\ell +2)(\ell -1){\,}$.  
We can now solve for ${\,}K_{1\ell m}{\,},{\,}k_{2\ell m}{\,}$ 
in terms of ${\,}q_{1\ell m}{\,}$ and its radial derivative 
$(\partial_{r}q_{1\ell m})$, giving
$$\eqalignno{k_{1\ell m}\; &=\; {1
    \over{2(\lambda\;r\;+\;3\;m)}}\Biggl[{{r\;e^{-a}(\partial_{r}q_{1\ell
    m})}\over {(\lambda + 1)}}\;+\;q_{1\ell m}\Biggr]&(7.51)\cr
k_{2\ell m}\;&=\;
    -\;{{e^{a}}\over{4(\lambda\;r\;+\;3\;m)}}\Biggl[r(\partial_{r}q_{1\ell
    m})+\;e^{a}\Bigl(1-{{3m}\over r}\Bigr)q_{1\ell
    m}\Biggr]&(7.52)\cr}$$
Further, $H^{RW}_{\ell m}$ and $K^{RW}_{\ell m}$ can also be written
in terms of $k_{1\ell m}$ and $k_{2\ell m}$.
\end{section}

\begin{section}{Classical action for even-parity metric perturbations}
As in the case of odd-parity gravitational perturbations, we can
exploit the uniqueness of the RW gauge to simplify the even-parity
action and to obtain a general gauge-invariant form for the
even-parity classical action $S^{(2)}_{\rm class}$.  In the RW gauge,
this is
$$\eqalign{S^{(2)}_{\rm class}\biggl[(h^{(+)}_{ij})_{\ell m}\biggr]\;
 & =\;{1 \over {32\pi}}\;\int_{\Sigma_F}\; d^{3}x
\;\sum_{\ell \ell' mm'}
\;\biggl(\pi^{(+)ij}\biggr)_{\ell m}\;
\biggl(h^{(+)}_{ij}\biggr)^{*}_{\ell' m'}\cr
&+{1\over {32\pi}}\;\sum_{\ell  m}
\;\int^{R\infty}_{0}\;dr\;e^{a/2}\biggl(H^{RW*}_{\ell m}
\;P^{RW}_{H 2\ell m}\;+\;2\;K^{RW*}_{\ell m}
P^{RW}_{K\ell m}\biggr)\mid_T\cr}\eqno(8.1)$$
\noindent
Again, we would like to put the action into the form
$\int{\,}dr{\;}\psi{\,}(\partial_{t}\psi){\,}$,  where $\psi$ is
gauge-invariant and obeys a wave equation.  Since $q_{1\ell m}$ is the
only unconstrained gauge-invariant even-parity quantity which involves
only perturbations of the intrinsic 3-geometry, one might expect that
Eq.(8.1) should reduce to the form
$$S^{(2)}_{\rm class}\biggl[\{q_{1\ell m}\}\biggr]\; =\;
{1 \over{32\pi}} 
\sum^{\infty}_{\ell=2}\;\sum^{\ell}_{m=-\ell}\;
\int^{R\infty}_{0}\;dr\;\biggl(\pi_{1\ell m}q^{*}_{1\ell
m}\;+\;(\partial_{r} 
Z_{\ell m})\biggr)\;\Bigl\arrowvert_{t=T},\eqno(8.2)$$
\noindent
for some variable $Z_{\ell m}$, where $\pi_{1\ell m}$ is the
gauge-invariant momentum conjugate to $q_{1\ell m}.$  This is in fact
the case.  First, make in Eq.(8.1) the substitutions (as mentioned at
the end of Sec.7) for $H^{RW}_{2\ell m}$ and $K^{RW}_{\ell m}$ in
terms of $k_{1\ell m}$ and $k_{2\ell m}{\;}$; then substitute
the expressions (7.51,52) for $k_{1\ell m}$ and $k_{2\ell m}$ in
terms of $q_{1\ell m}{\,}$.  After several integrations by parts, 
we arrive at an action of the form (8.2), with
$$\eqalignno{\pi_{1\ell
    m}\;&=\;{{r^{2}}\over{2\;(\lambda\;r\;+\;3\;m)}} \Biggl[\hat
    P^{RW}_{\ell m}\;-\;\biggl(1-{{3m}\over r}\biggr)\;P^{RW}_{H_{2}\ell
    m}\;e^{3a/2}\Biggr]\cr
&-{1\over
    2}\;\partial_{r}\Biggl[{{r^{3}}\over{(\lambda\;r\;+\;3\;m)}}\;\biggl({{\hat
    P^{RW}_{\ell m}\;e^{-a}}\over{(\lambda\;+\;1)}}\;
    -\;P^{RW}_{H_{2}\ell m}\;e^{a/2}\biggr)\Biggr]{\quad},&(8.3)\cr
Z_{\ell m}\;&=\; r^{3}\;e^{a/2}\;P^{RW}_{H_{2}\ell m}K^{RW}_{\ell
    m}\;+\;{{r^{3}\;q_{1\ell
    m}}\over{2\;(\lambda\;r\;+\;3\;m)}}\Biggl[{{\hat P^{RW}_{\ell
    m}\;e^{-a}}\over{(\lambda+1)}}\;-\;P^{RW}_{H_{2}\ell
    m}e^{a/2}\Biggr]{\quad},&(8.4)\cr 
\hat P^{RW}_{\ell m}\;&=\;e^{a/2}\biggl(2\;P^{RW}_{K\ell m}\;-\; 
2\;P^{RW}_{H_{2}\ell m}\;-\;r\Bigl(\partial_{r}P^{RW}_{H_{2}\ell
    m}\Bigl)\biggr){\quad}.&(8.5)\cr}$$ 

This expression for the even-parity classical action simplifies yet
further, since the linearised field equations imply that
$$\hat P^{RW}_{\ell m}\; =\; {{(\lambda +1)}\over r}\;e^{a}\;
H^{RW}_{1\ell m}.\eqno(8.6)$$
\noindent
Further, Eqs.(7.43,44) show, with the help of Eq.(7.48), that
$$\pi_{1\ell
  m}\;=\;{{\lambda\;e^{a}}\over{\bigl(2\;\lambda\;+\;{{6\;m}\over
  r}\bigr)}}\; (\partial_{t}q_{1\ell m}), \eqno(8.7)$$
\noindent
Eq.(8.2) for the even-parity ${\,}S^{(2)}_{\rm class}{\,}$ 
then reduces to an expression of the desired form:
$$\eqalign{S^{(2)}_{\rm class}\biggl[&\{(h^{(+)}_{ij})_{\ell m}\}\biggr]\cr 
&={\;}{\,}{{1}\over{32\pi}}\;\sum^{\infty}_{\ell =2}\;
\sum^{\ell}_{m=-\ell}\;{{(\ell -2)!}\over{(\ell +2)!}}\;
\int^{R\infty}_{0}\;dr\;e^{a}\xi^{(+)}_{2\ell m}\;
(\partial_{t}\xi^{(+)*}_{2\ell m})\Bigl\arrowvert_{t=T}{\quad},\cr}
\eqno(8.8)$$ 
\noindent
where ${\,}\xi^{(+)}_{2\ell m}{\,}$ is defined as
$$\xi^{(+)}_{2\ell m}\;
=\;{{\lambda\;q_{1\ell m}}\over{\bigl(\lambda\;+\;{{3m}\over r}\bigr)}}
{\quad}.\eqno(8.9)$$

We have used the property that the specified perturbations
$h^{(1)}_{ij}{\mid}_{\Sigma_F}$ of the spatial 3-metric on the final
boundary $\Sigma_F$ are taken to be real.  Of course, for the
Dirichlet boundary-value problem with $T$ rotated into the complex,
the classical solution for the metric and scalar field will have an
imaginary part and a real part.

Given the uniqueness of the RW gauge for even-parity modes, one can
see that Eq.(8.8) for $S^{(2)}_{\rm class}$ is in fact valid in any
gauge, with a vanishing contribution from the total divergence
as $q_{1\ell m}$ as given by Eq.(7.28), and therefore also
$\xi^{(+)}_{2\ell m}$, is gauge-invariant.  There are obvious
similarities between Eq.(8.8) and the classical massless-scalar action
of [3], with $\xi^{(+)}_{2\ell m}$ and $\xi_{0\ell m+}$
differing only by an $\ell$-dependent normalisation factor.  This
should not be so surprising, as scalar spherical harmonics have even
parity.

Again, one can show that the gauge-invariant quantity
${\,}\xi^{(+)}_{2\ell m}{\,}$ satifies Zerilli's equation [23]
$$e^{-a}\;\partial_{r}\bigl(e^{-a}\;(\partial_{r}\xi^{(+)}_{2\ell
m})\bigr)\;-(\partial_{t})^{2}\xi^{(+)}_{2\ell m}\;
-\;V^{(+)}_{\ell}\;\xi^{(+)}_{2\ell m}\; =\; 0{\quad},\eqno(8.10)$$
$$V^{(+)}_{\ell} =
\biggl(1-{{2m}\over r}\biggr)
{\Bigl({{2\lambda^{2}(\lambda +1)r^{3}+6\lambda^{2}mr^{2}+18\lambda m^{2}r
+18m^{3}}}\Bigr)
\over{r^{3}\;\bigl(\lambda{\,}r{\,}+{\,}3{\,}m\bigr)^{2}}}{\quad}>{\;}{\,}0
{\quad}.\eqno(8.11)$$
\noindent
Now, both for odd and even parity, the field equations for the metric
perturbations have been reduced to the two wave equations (4.12) and 
(8.10).
\end{section}

\begin{section}{Boundary conditions for even-parity perturbations}
In contrast to the odd-parity case, where we assumed an initially
stationary multipole, here for even parity we treat 
${\,}\xi^{(+)}_{2\ell m}{\,}$ by analogy with the massless-scalar-field 
quantity ${\,}\xi_{0\ell m+}{\,}$, and impose the Dirichlet boundary 
condition 
$$\xi^{(+)}_{2 \ell m}\;(0,r)\;=\;0\eqno(9.1)$$
at the initial surface $\Sigma_{I}{\;}(t=0)$.  Proceeding now by analogy
with the separation-of-variables analysis of Sec.5 for the odd-parity
case, we find that, if $K^{RW}_{\ell m}$ has $\sin(kt)$
time-dependence, then so must $H^{RW}_{\ell m}$ also, whereas
$H^{RW}_{1\ell m}$ must have $\cos(kt)$ time-dependence.  Consistency
with the gauge transformations (7.33-39) implies that these time
dependences are valid in an arbitrary gauge, and further that 
$G_{\ell m}$ and $h^{(+)}_{1\ell m}$ have $\sin(kt)$ time dependence, 
whereas $h^{(+)}_{0\ell m}$ has $\cos(kt)$ time dependence.  Consequently,
$q_{1\ell m}$ must have $\sin(kt)$ time dependence, whence the 
boundary condition (9.1) is justified through Eq.(8.9).
(Alternatively, one could instead have studied normal-mode 
time dependence.)

Following the scalar-field analysis of [3], we can write
$$\xi^{(+)}_{2 \ell m}(t,r)\;
=\;\int^{\infty}_{-\infty}\;dk{\;}{\,}a^{(+)}_{2k\ell m}\;
\xi^{(+)}_{2k\ell}(r){{\sin(kt)}\over{\sin(kT)}}{\quad},\eqno(9.2)$$
\noindent
where the $\{a^{(+)}_{2k\ell m}\}$ are suitable even-parity 'Fourier
coefficients', and where  $\{\xi^{(+)}_{2k\ell m}(r)\}$ are real
radial functions.  These functions satisfy
$$e^{-a}{d\over{dr}}\;\biggl(e^{-a}\;{{d\xi^{(+)}_{2k\ell}}\over{dr}}\biggr)
\;+\;\biggl[k^{2}\;-\;V^{(+)}_{\ell}(r)\biggr]\;\xi^{(+)}_{2k\ell}\;
=\; 0{\quad}.\eqno(9.3)$$
\noindent
Regularity at the origin implies that
$$\xi^{(+)}_{2k\ell}\;(r)\quad \sim \quad {\rm (const.)}\times{\;}
r{\;}j_{\ell}(kr)\eqno(9.4)$$ 
for small $r{\,}$.  Again, at large $r{\,}$, the potential vanishes
sufficiently rapidly that $\xi^{(+)}_{2k\ell}(r)$ has the asymptotic
form
$$\xi^{(+)}_{2k\ell}\;(r)\quad 
\sim\quad\biggl(\bigl(z^{(+)}_{2k\ell}\bigr)\;\exp(ikr^{*}_{s})\;
+\;\bigl(z^{(+)*}_{2k\ell}\bigr)\;\exp(-ikr^{*}_{s})\biggr){\quad},
\eqno(9.5)$$
\noindent
where $\{z^{(+)}_{2k\ell m}\}$ are complex constants.  Then the
classical action $S^{(2)}_{\rm class}$ for even-parity gravitational
perturbations reads
$$\eqalign{&S^{(2)}_{\rm class}\bigl[\{a^{(+)}_{2k\ell m}\}\bigr]\;\cr
  &=\;{1\over
  {16}}\; \sum^{\infty}_{\ell =2}\;\sum^{\ell}_{m=-\ell}\;{{(\ell
  -2)!}\over{(\ell +2)!}}\;\int^{\infty}_{0}dk\;k\;{\mid}
z_{2k\ell+}{\mid}^{2}\;{\mid}(a_{2k\ell m+})\; 
+\;(a_{2,-k\ell m+}){\mid}^{2}\;\cot(kT){\quad},\cr}\eqno(9.6)$$ 
\noindent
where the notation is in line with that for spin-0 and for 
odd-parity fields.  The coordinates  $\{a_{2k\ell m+}\}$ label the
configuration in $k$-space of the even-parity part of the metric 
perturbations on the final surface $\Sigma_F{\,}$.

Let us now re-assemble both the odd-and even-parity metric
perturbations.  As above, we consider for simplicity odd-parity metric
perturbations which are initially static (Neumann problem) and
even-parity metric perturbations which vanish initially (Dirichlet
problem), on the space-like hypersurface $\Sigma_I$.  The total
classical spin-2 action is then
$$\eqalign{&S^{(2)}_{\rm class}\;=\; {1 \over {32\pi}}\;\sum_{\ell mP}\;
{{(\ell -2)!}\over{(\ell +2)}}\;
P\;\int^{R\infty}_{0}\;dr\;e^{(a-b)/2}\;\xi_{2\ell
  mP}\;\Bigl(\partial_{t}\xi^{*}_{2\ell mP}\Bigr)
\Bigl\arrowvert_{\Sigma_F}\cr
&=\;{1 \over{16}}\; \sum^{\infty}_{\ell
  =2}\;\sum^{\ell}_{m=-\ell}\;\sum_{P=\pm} {{(\ell
  -2)!}\over{(\ell +2)!}}\;\int^{\infty}_{0}dk\;k\;{\mid}
z_{2k\ell P}{\mid}^{2}\;{\mid}(a_{2k\ell mP})
+\;(Pa_{2,-k\ell mP}){\mid}^{2}\;\cot(kT){\quad},\cr}\eqno(9.7)$$
\noindent
where the complex coefficients $\{a_{2k\ell mP}\}$ obey
$$a_{2k\ell mP}\; =\; P\;(-1)^{m}\; a^{*}_{2,-k\ell,-mP}{\quad}.
\eqno(9.8)$$
\noindent
Here, $P$ takes the value $\pm 1$ according as the parity 
is even or odd.

As in the case of odd-parity metric perturbations (Sec.6), the
even-parity metric perturbations also diverge at large $r{\,}$, 
except in a special gauge, the asymptotically-flat (AF) gauge.  
In the AF gauge for even parity, as for odd parity, all physical 
components
$h^{(1)}_{(\mu)(\nu)}
={\mid}\gamma^{\mu\mu}\gamma^{\nu\nu}{\mid}^{1\over 2}h^{(1)}_{\mu\nu}$ 
(that is, all components of $h^{(1)}_{\mu\nu}$ projected onto the legs 
of a pseudo-orthonormal tetrad oriented along the unperturbed
$(t,r,\theta,\phi)$ directions) fall off in the wave zone more rapidly
than $r^{-1}{\,}$, except for the transverse (angular) components, 
which carry information about the gravitational radiation.  
In the new (AF) gauge, for even parity, one has
$$h^{(+)AF}_{0\ell m}\; =\;H^{AF}_{0\ell m}\; =\;H^{AF}_{1\ell m}\;
=\; 0\eqno(9.9)$$ 
\noindent
Further, from the even-parity gauge transformations (7.23-29), 
one has
$$\eqalignno{0\;&=\;H^{RW}_{0\ell m}\;\;-\;a'\;\hat X^{(+)}_{1\ell
    m}\;+\;2\;\Bigl(\partial _{t}\hat X^{(+)}_{0\ell m}\Bigr),&(9.10)\cr
0\;&=\;H^{RW}_{1\ell m}\;+\;+\;e^{-a}\Bigl(\partial _{r}\hat X^{(+)}_{0\ell
    m}\Bigr)\;-\;e^{a}\Bigl(\partial _{t}\hat X^{(+)}_{1\ell m}\Bigr)
,&(9.11)\cr
H^{AF}_{2\ell m}\;&=\;H^{RW}_{2\ell m}
-\;a'\;\hat X^{(+)}_{1\ell
    m}\;-\;2\;\Bigl(\partial_{r}\hat X^{(+)}_{1\ell m}\Bigr),&(9.12)\cr
K^{AF}_{\ell m}\;&=\;K^{RW}_{\ell m}\;-\;{{2\;\hat X^{(+)}_{1\ell
    m}}\over r},&(9.13)\cr
G^{AF}_{\ell m}\;&=\;-\;2\;\hat X^{(+)}_{2\ell m},&(9.14)\cr
0\;&=\;e^{-a}\;\hat X^{(+)}_{0\ell m}\;-\;r^{2} 
\Bigl(\partial_{t}\hat X^{(+)}_{2\ell m}\Bigr),&(9.15)\cr
h^{(+)AF}_{1\ell m}\; &=\;-\;e^{a}\;\hat X^{(+)}_{1\ell m}\;-\;r^{2} 
\Bigl(\partial_{r}\hat X^{(+)}_{2\ell m}\Bigr),&(9.16)\cr}$$
\noindent
where a hat denotes a gauge function in the AF gauge.  Therefore, once
given $H^{RW}_{\ell m},\;H^{RW}_{1\ell m}{\,}$, 
and $K^{RW}_{\ell m}{\,}$, 
then Eqs.(9.10,11) can be solved for $\hat X^{(+)}_{0\ell m}$
and $\hat X^{(+)}_{1\ell m}{\;}$.  Thence, Eq.(9.15) can be used, 
in order to solve for $\hat X^{(+)}_{2\ell m}{\;}$.  
In solving these equations, one chooses the arbitrary functions 
which arise such that asymptotic flatness is still satisfied.  
Thus, the AF gauge is consistent.
\end{section}

\begin{section}{Boundary conditions in 2-spinor language}
In parallel with Sec.6 of [1], which was concerned with $s=1$
(Maxwell) perturbations, we describe here, in terms of 2-component
spinor language [8,9,26], the boundary conditions found in Secs.5,9
to be appropriate both for odd- and even-parity vacuum $s=2$
perturbations (gravitons).  We claim that in (near-) vacuum, the $s=2$
boundary data involve prescribing the magnetic part of the Weyl tensor
$C_{\alpha\beta\gamma\delta}$ [8-11] on $\Sigma_I$ and on 
$\Sigma_F{\,}$.  For simplicity, in the preceding sections, we took 
the initial data on $\Sigma_I$ to be zero.  Of course,{\it in vacuo}, 
where the Ricci tensor obeys $R_{\alpha\beta}{\,}={\,}0{\;}$, one has
$C_{\alpha\beta\gamma\delta}{\,}={\,}R_{\alpha\beta\gamma\delta}{\;}$, 
the Riemann tensor.

The algebraic symmetries of the Weyl tensor at a point 
are summarised by
$$\eqalign{C_{\alpha\beta\gamma\delta}&=\;
C_{[\alpha\beta][\gamma\delta]}\;=\;C_{\gamma\delta\alpha\beta},\cr
C_{\alpha[\beta\gamma\delta]}&= \;0
\;,\;\;C^{\alpha}_{~\beta\alpha\delta}\;= 0{\quad}.\cr}\eqno(10.1)$$

These imply that $C_{\alpha\beta\gamma\delta}$ has 10 algebraically
independent components at each point.  At a bounding space-like
hypersurface, such as $\Sigma_F{\,}$, let $n^{\mu}$ denote the unit
time-like (future-directed) normal vector to $\Sigma_F{\,}$.  
Then one can apply a $(3+1)$ decomposition to 
$C_{\alpha\beta\gamma\delta}$, which splits into two symmetric 
trace-free spatial tensors, the electric part $E_{ik}$ and the 
magnetic part $H_{ik}$ of the Weyl tensor [10,11].  Thus, 
the 10 space-time components of $C_{\alpha\beta\gamma\delta}$ 
have been decomposed into the 5 spatial components of $E_{ik}$ 
and 5 more of $H_{ik}$.  (Correspondingly, in Maxwell theory, 
the 6 non-trivial components of the field strength $F_{\mu\nu}$ 
become the 3 of $E_i$ plus the 3 of $B_i{\,}$.)

For convenience of exposition, consider an 'adapted' coordinate system
$(x^0,x^1,x^2,x^3)$ in a neighbourhood of $\Sigma_F{\,}$, such that
$\Sigma_F$ lies at $x^{0}{\,}={\,}0{\,}$, and such that $n^{0}{\,}=1$ 
at all points of $\Sigma_F{\,}$.  The spatial 3-metric is, as usual, 
denoted by $h_{ij}{\,}$, and we write 
${\,}h{\,}={\,}{\rm det}(h_{ij}){\,}$.  Then the electric part of the 
Weyl tensor is defined to be
$$E_{\alpha\gamma}\; 
=\; C_{\alpha\beta\gamma\delta}\; n^{\beta}\; n^{\delta}{\quad}.\eqno(10.2)$$
\noindent
In an adapted coordinate system, this reads
$$E_{ik}\; =\;C_{i0k0}{\quad}.\eqno(10.3)$$
\noindent
The magnetic part of the Weyl tensor is defined to be
$$H_{\alpha\gamma}\; 
=\; {{1}\over{2}}\;\eta_{\alpha\beta}^{~~~\rho\sigma}\;
C_{\rho\sigma\gamma\delta}\; n^{\rho}\; n^{\delta}{\quad},\eqno(10.4)$$
\noindent
where
$$\eta_{\alpha\beta\gamma\delta}{\;}{\,}
={\;}{\,}n_{[\alpha\beta\gamma\delta]}{\;}{\,}
={\;}{\,}(-g)^{{1}\over{2}}{\;}{\,}\epsilon_{\alpha\beta\gamma\delta}
\eqno(10.5)$$
\noindent
is the alternating tensor, with 
${\,}g{\,}={\,}{\rm det}(g_{\mu\nu}){\,}$ 
and 
$\epsilon_{\alpha\beta\gamma\delta}\; 
=\;\epsilon_{[\alpha\beta\gamma\delta]}$ the alternating symbol,
normalised such that ${\,}\epsilon_{0123}{\,}={\,}0{\,}$.  
In an adapted coordinate system, one finds that
$$H_{ik}\; =\; -{\,}{{1}\over{2}}{\;}h^{-1/2}{\;}{\,}h_{in}{\;}{\,}
\epsilon^{n\ell m}{\;}C_{\ell m k 0}{\quad}.\eqno(10.6)$$
\noindent
Both $E_{ik}$ and $H_{ik}{\,}$, so defined, are the components of
3-dimensional (spatial) tensors, obeying the algebraic restrictions
(symmetric, traceless):
$$\eqalignno{E_{ik} &=\; E_{(ik)}\; ,
{\qquad}h^{ik}\;E_{ik}\; = 0{\quad};&(10.7)\cr
H_{ik}&=\; H_{(ik)}\; ,
{\qquad}h^{ik}\; H_{ik}\; = 0{\quad}.&(10.8)\cr}$$
\noindent
By analogy with the vacuum Maxwell case for $E_i{\,},{\,}B_i{\,}$ 
of [1], here $E_{ik}$ and $H_{ik}$ also obey differential
constraints on the bounding 3-surface.  From the Bianchi identities 
[8,9]  one has ({\it in vacuo})
$$^{3}\nabla_{k}E^{ik}\; =\;0,{\qquad}^{3}\nabla_{k}H^{ik}\; 
=\; 0{\quad},\eqno(10.9)$$ 
\noindent
where $^{3}\nabla_k$ denotes an intrinsic 3-dimensional covariant
derivative, which preserves the 3-metric $h_{ij}{\;}$.

In 2-component spinor language [8,9], one has 
$$C_{\alpha\beta\gamma\delta}{\quad}
\leftrightarrow{\quad}\epsilon_{AB}\;\epsilon_{CD} \;
\tilde\Psi_{A'B'C'D'}\;
+\;\epsilon_{A'B'}\;\epsilon_{C'D'}\; \Psi_{ABCD}{\quad},\eqno(10.10)$$
\noindent
where
$$\Psi_{ABCD}{\;}{\,}={\;}{\,}\Psi_{(ABCD)}\eqno(10.11)$$
\noindent
is the totally symmetric (complex) Weyl spinor, 
and (in Lorentzian signature) $\tilde\Psi_{A'B'C'D'}$ 
is its hermitian conjugate.  The dual of the Weyl tensor is defined as
$$^{*}C_{\alpha\beta\gamma\delta}\;
=\;{1\over2}\;\eta_{\alpha\beta\rho\sigma}\;
C^{\rho\sigma}_{~~~\gamma\delta}{\quad}.\eqno(10.12)$$  
\noindent
One finds (cf. Sec.6 of [1]) that:
$$\Bigl(C_{\alpha\beta\gamma\delta}\; 
+\; i\;^{*}C_{\alpha\beta\gamma\delta}\Bigr){\quad}
\leftrightarrow{\quad}2\;\epsilon_{A'B'}\;\epsilon_{C'D'}\;
\Psi_{ABCD}{\quad}.\eqno(10.13)$$
\noindent
The (vacuum) Bianchi identities read (again, cf. Sec.6 of [1]):
$$\nabla^{AA'}\;\Psi_{ABCD}\;=\; 0{\quad}.\eqno(10.14)$$
\noindent
On the bounding surface $\Sigma_F$ (say), one finds analogously that 
$$E_{k\ell}\; + i\; H_{k\ell}\; =
2\;\Psi_{ABCD}\;\bigl(n^{A}_{~~B'}\;e^{BB'}_{~~~~k}
\bigr)\bigl(n^{C}_{~~D'}\; e^{DD'}_{~~~~\ell}\bigr)\eqno(10.15)$$
\noindent
and its hermitian conjugate.  Thence, the magnetic tensor 
$H_{k\ell}$ is given by
$$H_{k\ell}\; =\;\biggl(-i\;\Psi_{ABCD}\;\Bigl(n^{A}_{~~B'}\;e^{BB'}_{~~~k}
\Bigr)\Bigl(n^{C}_{~~D'}\; e^{DD'}_{~~\ell}\Bigr)\biggr)\;+\;[{\rm
h.c.}]{\quad}.\eqno(10.16)$$
\noindent
Eq.(10.16) and its hermitian conjugate can straightforwardly be inverted
to give an expression analogous to $\Psi^{AB}_{+}{\,}$, as given in
Sec.6 of [1], 
which provides a spinorial version of the quantity $H_{k\ell}$ 
to be fixed on $\Sigma_F{\,}$.  Of course, the perturbative boundary 
data $H_{k\ell}$ must be chosen such that the divergence condition 
(10.9) holds: $^{3}\nabla_{k}H^{ik}{\,}={\,}0$ on $\Sigma_F{\,}$, 
just as, in Sec.6 of [1], the condition
$^{3}\nabla_{k}B^{k}{\,}={\,}0$ must hold.

Two of the five complex components of ${\,}\Psi_{ABCD}{\,}$ 
are contained in the Newman-Penrose quantities [8,9,27]
$$\Psi_{0}\; =\;\Psi_{ABCD}\;
o^{A}\;o^{B}\;o^{C}\;o^{D}{\quad},
{\qquad}\Psi_{4} \;=\;\Psi_{ABCD}\;
\iota^{A}\;\iota^{B}\; \iota^{C}\;\iota^{D}{\quad},\eqno(10.17)$$
\noindent
where $(o^{A}{\,},\iota^{A})$ is a normalised spinor dyad [1].
Such a dyad $(o^{A},{\,}\iota^{A})$ is a pair of spinors which
provide a basis for the 2-complex-dimensional vector space of 
spinors $\omega^A$ at each point, and is normalised according to
$o_{A}{\;}\iota^{A}{\;} 
={\;}1{\;} 
={\;}-{\,}\iota_{A}{\;}o^{A}{\;}$. 
Knowledge of a null tetrad [27] 
${\,}\ell^{\mu}{\,},{\,}n^{\mu}{\,},{\,}m^{\mu}{\,},{\,}\bar m^{\mu}$ 
of vectors at a point is equivalent to knowledge of the corresponding 
normalised spinor dyad $(o^{A},\;\iota^{A})$, through the relations
$l^{\mu}{\;}\leftrightarrow{\;}o^{A}{\;}o^{A'}{\;}, 
{\;}n^{\mu}{\;}\leftrightarrow{\;}\iota^{A}{\;}\iota^{A'}{\;},
m^{\mu}{\;}\leftrightarrow{\;}o^{A}{\;}\iota^{A'}{\;},
{\;}\bar m^{\mu}{\;}\leftrightarrow{\;}\iota^{A}{\;}o^{A'}{\;}$.

In the case of a spherically-symmetric black-hole background 
(or indeed of a rotating Kerr black hole), taking the Kinnersley 
null tetrad [28,29], which corresponds to a particular choice of 
$(o^{A},\iota^{A})$, it was shown by Teukolsky [30] that
$\Psi_{0}$ (and ${\,}\Psi_{4}$) each obey decoupled separable wave
equations.  Following work by Chrzanowski [31], it was confirmed 
by Wald [32] that, given a solution $\Psi_0$ (or $\Psi_4$) of the
Teukolsky equation for a (nearly-) Kerr background, all the vacuum
metric perturbations can be reconstructed in a certain gauge through 
a series of simple operations on $\Psi_0$ (or $\Psi_4$) [29,31,32].
Once the linearised metric perturbations are known, then, of course,
one can compute other Newman-Penrose quantities, such as
$$\Psi_{2}{\;}{\,}
={\;}{\,}\Psi_{ABCD}\;o^{A}\;o^{B}\;\iota^{C}\;\iota^{D}{\quad}.
\eqno(10.18)$$

The analogous process can also be carried out for spin-1 (Maxwell)
perturbations of a spherically-symmetric black-hole solution 
or of the Kerr metric [29,31,32].  For $s=1$ perturbations 
of the Kerr metric, as mentioned in Sec.6 of [1], the Newman-Penrose 
quantities $\phi_{0}{\,},{\,}\phi_{2}$ each obey a decoupled separable 
equation; then [29,31,32] the corresponding linearised Maxwell vector 
potential $A_{\mu}$ (in a particular gauge) can be reconstructed 
by simple steps from $\phi_0$ (or from $\phi_2$).  
Hence, the middle Newman-Penrose quantity $\phi_1$ can also be found; 
it is $\phi_1$ which admits the expansion
$$\phi_{1}\;=\; {1 \over {2 r^{2}}}\;\sum_{\ell m}\;
\Bigl(\psi_{1\ell m}^{(e)}\;
+i\;\psi_{1\ell m}^{(o)}\Bigr)\;Y_{\ell m}(\Omega)\eqno(10.19)$$  
\noindent
given in Eq.(6.22) of [1], which then allows comparison with the 
language used in Secs.2-5 of [1].  Although we have not quite 
finished detailed calculations on this point, it does now look 
reasonable to expect that, for $s=2$ gravitational perturbations, 
there should exist a relation analogous to Eq.(10.19).
\end{section}

\begin{section}{Conclusion}
Linearised gravitational-wave $(s=2)$ perturbations about a
spherically-symmetric Einstein/massless-scalar collapse to a black
hole have been studied in this paper.  In the companion paper [1], 
for Maxwell $(s=1)$ perturbations, the principal aims were 
(1) to specify suitable perturbative boundary data on the final 
space-like hypersurface $\Sigma_F$ at a late time $T$, subject 
(for simplicity) to the initial boundary data on $\Sigma_I{\,}$ 
(time $t=0{\,}$) being spherically symmetric.  
(2) To express the Lorentzian classical action $S_{\rm class}$ 
as an explicit functional of the 'suitable' boundary data above, 
and of the proper-time interval $T{\,}$, once $T$ has been rotated 
into the complex: 
$T\rightarrow{\mid}T{\mid}\exp(-i\theta)$, 
for $0<\theta\leq\pi/2{\,}$.  
(3)  Given $S_{\rm class}{\;}$, to compute, following Feynman, 
the quantum amplitude for the weak-field final data, by taking 
the limit of the semi-classical amplitude 
${\rm (const.)}\times\exp(iS_{\rm class})$ 
as $\theta\rightarrow 0_{+}{\,}$.

The same aims hold for the present paper, for gravitational-wave
perturbations $(s=2)$, with analogous results.  As in the $s=1$ case,
it is also necessary here to decompose the metric perturbations into 
parts with odd and even parity.  The main difference on moving from 
$s=1$ [1] to the present paper is a considerable increase in algebraic 
or analytic complexity, to be expected since one deals with tensor 
fields rather than vector fields.

Some indications towards unification of these ideas for perturbative
fields of different spin $s$ are contained in Sec.10, which parallels
for $s=2$ the treatment of $s=1$ in [1].  For $s=1{\,}$, the
quantity most naturally specified as an argument of the quantum
wave-functional, on a bounding hypersurface such as $\Sigma_F{\,}$, 
is the (spatial) magnetic field $B_i{\,}$, subject to the condition
$^{3}\nabla_{k}B^{k}{\,}={\,}0{\;}$.  Correspondingly, for linearised
gravitational waves $(s=2)$, the natural boundary data were found to be
the (symmetric, trace-free) magnetic part $H_{ik}$ of the Weyl tensor
[10,11], subject to $^{3}\nabla_{k}H^{ik}{\,}={\,}0{\;}$.  
In 2-component spinor language, these correspond $(s=1)$ 
to a particular 'projection' of the (complex) symmetric Maxwell 
field-strength spinor $\phi_{AB}{\,}={\,}\phi_{(AB)}{\,}$, 
and $(s=2)$ to a projection of the totally-symmetric (complex) Weyl spinor 
$\Psi_{ABCD}{\,}={\,}\Psi_{(ABCD)}{\,}$. Of course, as treated in
[1,33], these boundary conditions constructed from
$\phi_{AB}{\,}$ and $\Psi_{ABCD}{\,}$ are special cases of the natural
boundary conditions for gauged supergravity [34-36].  
Although 2-component spinor 
language might seem a luxury in treating bosonic fields describing 
photons or gravitons, above, it is practically a necessity in treating 
the corresponding fermionic (massless) neutrino spin-${{1}\over{2}}$ 
field, as in [33], and (for supergravity) the gravitino spin-$3 \over
2$ field, on which work is in progress [37].
\end{section}

\parindent = 1 pt

\begin{section}*{References}
\everypar{\hangindent\parindent}

\noindent [1] A.N.St.J.Farley and P.D.D'Eath, 
'Spin-1 Amplitudes in Black-Hole Evaporation', 
submitted for publication (2005).

\noindent [2] A.N.St.J.Farley and P.D.D'Eath, 
'Quantum Amplitudes in Black-Hole Evaporation: I. Complex Approach', 
submitted for publication (2005);
Phys Lett. B, {\bf 601}, 184 (2004).        

\noindent [3] A.N.St.J.Farley and P.D.D'Eath, 
'Quantum Amplitudes in Black-Hole Evaporation: II. Spin-0 Amplitude', 
submitted for publication (2005).

\noindent [4] R.P.Feynman and A.R.Hibbs, 
{\it Quantum Mechanics and Path Integrals}, 
(McGraw-Hill, New York) (1965).

\noindent [5] J.Mathews, J. Soc. Ind. Appl. Math. {\bf 10}, 768 (1962).

\noindent [6] J.N.Goldberg, A.J.MacFarlane, E.T.Newman, F.Rohrlich and 
E.C.G.Sudarshan, J. Math. Phys. {\bf 8}, 2155 (1967).

\noindent [7] T.Regge and J.A.Wheeler, Phys. Rev. {\bf 108}, 1063 (1957).

\noindent [8] R.Penrose and W.Rindler, {\it Spinors and Space-Time}, vol. 1
(Cambridge University Press, Cambridge) (1984).

\noindent [9] R.Penrose and W.Rindler, {\it Spinors and Space-Time}, vol. 2
(Cambridge University Press, Cambridge) (1986).

\noindent [10] P.D.D'Eath, Phys. Rev. D {\bf 11}, 1387 (1975)

\noindent [11] K.S.Thorne, R.H.Price and D.A.Macdonald, 
{\it Black holes. The membrane paradigm.} 
(Yale University Press, New Haven, Ct.) (1986).

\noindent [12] P.D.D'Eath, 
'What local supersymmetry can do for quantum cosmology', in 
{\it The Future of Theoretical Physics and Cosmology}, 
eds. G.W.Gibbons, E.P.S.Shellard and S.J.Rankin 
(Cambridge University Press, Cambridge) 693 (2003).

\noindent [13] P.D.D'Eath, 
'Loop amplitudes in supergravity by canonical quantization', 
in {\it Fundamental Problems in Classical, Quantum 
and String Gravity}, 
ed. N.S\' anchez (Observatoire de Paris) 
{\bf 166} (1999), hep-th/9807028.

\noindent [14] K.S.Thorne and A.Campolattaro, 
Astrophys. J. {\bf 149}, 591 (1967).

\noindent [15] V.Moncrief, Ann. Phys. (N.Y.) {\bf 88}, 323 (1974).

\noindent [16]  C.W.Misner, K.S.Thorne and J.A.Wheeler, {\it Gravitation},
(Freeman, San Francisco) (1973).

\noindent [17] A.N.St.J.Farley and P.D.D'Eath, 
'Vaidya Space-Time in Black-Hole Evaporation', 
submitted for publication (2005).

\noindent [18] P.C.Vaidya, Proc. Indian Acad. Sci. {\bf A33}, 264 (1951).

\noindent [19]  R.W.Lindquist, R.A.Schwartz and C.W.Misner, 
Phys Rev. {\bf 137}, 1364 (1965).

\noindent [20] D.N.Page, Phys. Rev. D {\bf 13}, 198 (1976).

\noindent [21] A.N.St.J.Farley and P.D.D'Eath, Phys. Lett. B {\bf
613}, 181 (2005). 

\noindent [22] F.J.Zerilli, Phys. Rev. D {\bf 2}, 2141 (1970).

\noindent [23] F.J.Zerilli, Phys. Rev. D {\bf 9}, 860 (1974).

\noindent [24] C.T.Cunningham, R.H.Price and V.Moncrief, 
Astrophys. J. {\bf 224}, 643 (1978); ibid. {\bf 230}, 870 (1979).

\noindent [25] S.Chandrasekhar, {\it The Mathematical Theory of Black Holes}
(Oxford University Press, Oxford) (1992).

\noindent [26] P.D.D'Eath, {\it Supersymmetric Quantum Cosmology}, 
(Cambridge University Press, Cambridge) (1996).

\noindent [27] E.T.Newman and R.Penrose, J. Math. Phys. {\bf 3}, 566 (1962).

\noindent [28] W.Kinnersley, J. Math. Phys {\bf 10}, 1195 (1969).

\noindent [29] J.A.H.Futterman, F.A.Handler and R.A.Matzner,   
{\it Scattering from Black Holes} 
(Cambridge University Press, Cambridge) (1988).

\noindent [30] S.A.Teukolsky, Astrophys. J. {\bf 185}, 635 (1973).

\noindent [31] P.L.Chrzanowski, Phys. Rev D {\bf 11}, 2042 (1975).

\noindent [32] R.M.Wald, Phys. Rev. Lett. {\bf 41}, 203 (1978).

\noindent [33] A.N.St.J.Farley and P.D.D'Eath, Class. Quantum
Grav. {\bf 22}, 3001 (2005).

\noindent [34] P.Breitenlohner and D.Z.Freedman, 
Phys. Lett. B {\bf 115}, 197 (1982).

\noindent [35] P.Breitenlohner and D.Z.Freedman, 
Ann. Phys. (N.Y.) {\bf 144}, 249 (1982).

\noindent [36] S.W.Hawking, Phys. Lett. B {\bf 126}, 175 (1983).

\noindent [37]  A.N.St.J.Farley and P.D.D'Eath, 
'Spin-3/2 Amplitudes in Black-Hole Evaporation', in progress.

\end{section}

\end{document}